\documentclass[aps,prd,preprintnumbers,groupedaddress,nofootinbib,eqsecnum,notitlepage,10pt]{revtex4-1}
\usepackage{here}
\usepackage{graphicx}
\usepackage{amsmath,amsthm,amssymb}
\usepackage{bm}

\usepackage{amsfonts}
\usepackage{dcolumn}
\usepackage{hyperref}
\allowdisplaybreaks[1]
\begin{document}

\newcommand{\newc}{\newcommand}
\newc{\tif}{\tilde{f}}
\newc{\tih}{\tilde{h}}
\newc{\tip}{\tilde{\phi}}
\newc{\tiA}{\tilde{A}}
\newcommand{\ben}{\begin{eqnarray}}
\newcommand{\een}{\end{eqnarray}}
\newc{\be}{\begin{equation}}
\newc{\ee}{\end{equation}}
\newc{\ba}{\begin{eqnarray}}
\newc{\ea}{\end{eqnarray}}
\newc{\bea}{\begin{eqnarray*}}
\newc{\eea}{\end{eqnarray*}}
\newc{\D}{\partial}
\newc{\ie}{{\it i.e.} }
\newc{\eg}{{\it e.g.} }
\newc{\etc}{{\it etc.} }
\newc{\etal}{{\it et al.}}
\newcommand{\nn}{\nonumber}
\newc{\ra}{\rightarrow}
\newc{\lra}{\leftrightarrow}
\newc{\lsim}{\buildrel{<}\over{\sim}}
\newc{\gsim}{\buildrel{>}\over{\sim}}

\makeatother

\preprint{YITP-18-117}

\title{
Constraints on massive vector dark energy models \\
from integrated Sachs-Wolfe-galaxy cross-correlations
}

\author{
Shintaro Nakamura$^{1}$, Antonio De Felice$^{2}$, 
Ryotaro Kase$^{1}$, and
Shinji Tsujikawa$^{1}$
}

\vspace{1cm}

\affiliation{
$^1$Department of Physics, Faculty of Science, 
Tokyo University of Science, 1-3, Kagurazaka,
Shinjuku-ku, Tokyo 162-8601, Japan\\
$^2$Center for Gravitational Physics, Yukawa Institute for 
Theoretical Physics, 
Kyoto University, 606-8502, Kyoto, Japan 
}

\begin{abstract}
The gravitational-wave event GW170817,
together with the electromagnetic counterpart, shows that 
the speed of tensor perturbations $c_T$ on the cosmological 
background is very close to that of light $c$ for the redshift $z<0.009$. 
In generalized Proca theories, the Lagrangians 
compatible with the condition $c_T=c$ are constrained 
to be derivative interactions up to cubic order, besides 
those corresponding to intrinsic vector modes. 
We place observational constraints on a dark energy model 
in cubic-order generalized Proca theories with intrinsic vector modes 
by running the Markov chain Monte Carlo (MCMC) code. 
We use the cross-correlation data of the integrated 
Sachs-Wolfe (ISW) signal and galaxy distributions
in addition to the data sets of cosmic microwave 
background, baryon acoustic oscillations, type Ia supernovae,  
local measurements of the Hubble expansion rate, and 
redshift-space distortions. We show that, unlike cubic-order 
scalar-tensor theories, the existence of intrinsic vector modes 
allows the possibility for evading the ISW-galaxy anticorrelation 
incompatible with the current observational data. 
As a result, we find that the dark energy model in cubic-order 
generalized Proca theories exhibits a better fit to the data 
than the cosmological constant,
even by including the ISW-galaxy correlation data in the MCMC analysis.

\end{abstract}

\date{\today}

\maketitle

\section{Introduction}
\label{introsec}

The late-time cosmic acceleration has been probed by several different observations
such as the supernovae type Ia (SN Ia) \cite{Riess:1998cb, Perlmutter:1998np,Suzuki:2011hu}, 
the cosmic microwave background (CMB) \cite{WMAP2003,WMAP2012,Planck2013,Planck2015_14, Planck2018}, 
and baryon acoustic oscillations (BAOs) \cite{Eisenstein:2005su, 6dFGS, SDSS-MGS, BOSS, BOSS-CMASS, Wiggle-Z}. 
The simplest theoretical candidate for the source of this phenomenon 
is the cosmological constant $\Lambda$ \cite{Weinberg,Martin}, but it is generally difficult 
to explain the tiny observed dark energy scale from the vacuum energy 
arising in particle physics. From the observational viewpoint, 
there have been tensions for today's Hubble parameter 
$H_0$ constrained from the CMB \cite{Planck2013,Planck2018} 
and its direct measurements at low redshifts \cite{Riess:2016jrr}. 
It is worth pursuing alternative theoretical candidates for dark energy 
and studying whether they are a better fit to the data over 
the cosmological constant (see 
Refs.~\cite{dreview1,dreview2,dreview3,dreview4,dreview5,dreview6} for reviews).

There are dark energy models based on a scalar field $\varphi$ with 
a potential (quintessence \cite{quin1,quin2,quin3,quin4,quin5,quin6,quin7}) 
or with a nonlinear kinetic term ($k$-essence \cite{kes1,kes2,kes3}). 
As long as the ghost is absent, these models lead to the dark energy 
equation of state $w_{\rm DE}$ larger than $-1$.
In the presence of nonlinear scalar self-interactions and nonminimal/derivative 
couplings to gravity, Horndeski theories \cite{Horndeski} are the most
general scalar-tensor theories with second-order equations of 
motion \cite{Deffayet:2011gz, Kobayashi:2011nu, Charmousis:2011bf}. 
The subclass of Horndeski theories consists of Brans-Dicke 
theory \cite{Brans,Yoko,Umilta:2015cta,Ballardini:2016cvy}, 
$f(R)$ gravity \cite{fR1,fR2,fR3,fR4}, 
kinetic braidings \cite{braiding}, Galileons \cite{Gali1,Gali2} and so on, 
in which case it is possible to realize $w_{\rm DE}<-1$ 
with neither ghost nor
Laplacian instabilities \cite{Yoko,fR1,fR4,DT10}. 
Dark energy models in Horndeski theories can be distinguished 
from each other by the different evolution of $w_{\rm DE}$ as well as 
the different cosmic growth history \cite{DKT11,Amen12,Raveri14,Bellini,Zuma17}. 

A massive vector field with a broken $U(1)$ gauge symmetry can also 
be the source for dark energy.
The $U(1)$-broken vector-tensor theories with second-order equations 
of motion, which are known as generalized Proca (GP) 
theories \cite{Heisenberg,Tasinato1,Tasinato2,Allys1,Jimenez2016,Allys2}, 
contain vector nonlinear self-interactions and derivative couplings to gravity.
At the background level, the existence of a temporal vector component $A^0$ 
gives rise to a self-accelerating de Sitter attractor 
($w_{\rm DE}=-1$) \cite{cosmoGP}. 
The evolution of $w_{\rm DE}$ before approaching the attractor 
is different depending on the forms of derivative interactions.
There exist dark energy models in which $w_{\rm DE}$ is less than $-1$ during radiation and matter eras without theoretical inconsistencies. 
The presence of intrinsic vector modes provides an interesting 
possibility for realizing the gravitational interaction smaller than that in 
general relativity (GR) on scales relevant to the growth 
of large-scale structures \cite{GeffGP}.
In the solar system, the nonlinear cubic and quartic 
interactions can suppress the propagation of fifth forces under 
the operation of the Vainshtein mechanism \cite{Vain1,Vain2}.

The recent detection of the gravitational-wave event GW170817 \cite{GW170817} 
from a neutron star merger, together with the gamma-ray burst 
GRB170817A \cite{GRB170817A}, constrains the speed of gravity 
$c_T$ to be very close to that of light $c$ \cite{GWGRB}. 
If we strictly demand that $c_T=c$ in Horndeski theories, the quartic 
derivative and quintic interactions are not 
allowed \cite{GWcon1,GWcon2,GWcon3,GWcon6,GWcon4,GWcon5}. 
As for GP theories with a vector field $A^{\mu}$, the Lagrangian 
is restricted to be of the form ${\cal L}=G_2(X)+G_3(X) \nabla_{\mu}A^{\mu}+
(M_{\rm pl}^2/2)R$ with intrinsic vector modes, where $G_2, G_3$ are 
functions of $X=-A_{\mu}A^{\mu}/2$, $\nabla_{\mu}$ is 
the covariant derivative operator, $M_{\rm pl}$ is the reduced 
Planck mass, and $R$ is the Ricci scalar. 
Even with this restriction, the dark energy model proposed 
in Ref.~\cite{cosmoGP} is still cosmologically viable in that 
there exists a theoretically consistent parameter space 
with neither ghost nor Laplacian instabilities.

In Ref.~\cite{dFHT2017}, the MCMC likelihood analysis 
was performed for 
the dark energy model in full GP theories proposed in Refs.~\cite{cosmoGP,GeffGP} 
by exploiting the observational data of SNIa, CMB, BAOs, the Hubble 
expansion rate $H(z)$, and redshift-space distortions (RSDs).
In this model, the dark energy equation of state during the matter era can be 
expressed as $w_{\rm DE}=-1-s$, where $s$ is a positive constant. 
{}From the MCMC analysis, the parameter $s$ is constrained to be in 
the range $s=0.16^{+0.08}_{-0.08}$ (95 \%\,CL), so the model with $s>0$ 
is favored over the $\Lambda$-cold-dark-matter ($\Lambda$CDM) model ($s=0$). 
This is mostly attributed to the fact that the existence of the additional 
parameter $s$ can reduce the tension of $H_0$ between the measurements 
at high and low redshifts.

Since the dark energy dynamics is quantified by the single parameter $s$, 
the background cosmology in cubic-order GP theories is the same as 
that in full GP theories for a given value of $s$. 
However, the evolution of cosmological perturbations is affected by 
the quartic and quintic derivative couplings, so the cubic-order GP theories predict 
different cosmic growth histories from those in full GP theories. 
In cubic-order GP theories, the gravitational interaction for linear perturbations 
is always enhanced compared to that in GR \cite{Amen17}, 
while this is not generally the case in full GP theories \cite{GeffGP}. 
Moreover, the number of free parameters associated with evolution of 
perturbations is reduced by the absence of  quartic and quintic derivative couplings.
Then, at the level of perturbations, the MCMC analysis 
can lead to different constraints on cubic-order GP 
theories relative to those found in Ref.~\cite{dFHT2017}.

The cross-correlation between the ISW signal in CMB and galaxy distributions is another distinguished observable for probing the modification of 
gravity at large distances \cite{CTurok,Afsh,Cora,Pogosian,ISWdata,Gian,Ballardini:2017xnt}.
In dark energy models within the framework of GR, the ISW-galaxy 
cross-correlation is positive at any redshift. 
On the other hand, the cross-correlation can be negative for modified  
gravitational theories in which the normalized effective gravitational 
coupling $\Sigma$ associated with the light bending 
rapidly increases at low redshifts. Indeed, this negative cross-correlation can arise for 
cubic-order Horndeski theories like Galileons and 
their extensions \cite{Kobayashi:2009wr,Kimura:2011td}. 
The models in which scalar derivative cubic couplings 
play the dominant role in the dark energy dynamics 
can be ruled out from the ISW-galaxy cross-correlation 
data \cite{Renk}.

Taking the scalar limit $A^{\mu} \to \nabla^{\mu}\varphi$ for the coupling 
$G_3(X) \nabla_{\mu}A^{\mu}$ in GP theories, it recovers the shift-symmetric cubic 
Lagrangian in Horndeski theories. 
Then, one may wonder if the tendency of negative ISW-galaxy cross-correlations 
in cubic-order Horndeski theories also persists in cubic-order GP theories.
In GP theories, however, there exist intrinsic vector modes 
besides the longitudinal scalar mode. 
The former affects the evolution 
of scalar perturbations through the quantity $q_V$ 
relevant to the no-ghost condition 
of vector perturbations. Then, the observational predictions of cosmic growth rates 
and ISW-galaxy cross-correlations are generally different from those in 
cubic-order Horndeski theories.

In this paper, we place observational constraints on a dark energy model in 
cubic-order GP theories with intrinsic vector modes satisfying the condition $c_T=c$. 
In addition to the observational data of SNIa, CMB, BAOs, $H(z)$, and RSDs, 
we take into account the ISW-galaxy cross-correlation data in the MCMC analysis and study whether the model 
is subject to a tighter constraint than that derived previously.
We show that, even with the ISW-galaxy cross-correlation data, the models with $s>0$ 
are still favored over the $\Lambda$CDM model 
due to the existence of intrinsic modes.

This paper is organized as follows. 
In Sec.~\ref{sec2}, we briefly review the background cosmological dynamics and 
the evolution of perturbations for dark energy models in cubic-order GP theories.
In Sec.~\ref{sec3}, we discuss general conditions for generating the anticorrelated
ISW-galaxy spectrum and study the cases in which the negative ISW-galaxy 
correlation arises in GP theories.
In Sec.~\ref{sec4}, we perform the MCMC analysis by using the data mentioned above 
and put observational constraints on the model parameters. 
Section \ref{sec5} is devoted to conclusions.
In what follows, we use the units where the speed of light $c$ and the reduced Planck constant $\hbar$ are
equivalent to 1.

\section{Dark energy in GP theories}
\label{sec2}

We begin with the cubic-order GP theories given 
by the action \cite{Heisenberg,Tasinato1}
\be
{\cal S}= \int d^{4}x \sqrt{-g} \left[ G_2 \left( X,F,Y \right)
+G_3(X) \nabla_{\mu}A^{\mu}
+\frac{M_{\rm pl}^2}{2} R \right]
+{\cal S}_M\,,
\label{action}
\ee
where $g$ is the determinant of the metric tensor 
$g_{\mu\nu}$,
$G_2$ is a function of $X=-A_{\mu}A^{\mu}/2$, and 
\be
F=-\frac{1}{4} F_{\mu \nu} F^{\mu \nu}\,,\qquad
Y= A^{\mu}A^{\nu} {F_{\mu}}^{\alpha} 
F_{\nu \alpha}\,,
\ee
with $F_{\mu \nu}=\nabla_{\mu}A_{\nu}-\nabla_{\nu}A_{\mu}$. 
The cubic coupling $G_3$ is a function of $X$ alone. 
For the matter action ${\cal S}_M$, we consider perfect fluids 
minimally coupled to gravity. 
The quartic and quintic couplings in GP theories generally lead to 
$c_T^2$ different from unity \cite{cosmoGP}, so we do not consider
such contributions. The quintic intrinsic vector mode and 
the sixth-order nonminimal coupling \cite{Jimenez2016} can be 
added to the action (\ref{action}) without modifying 
the value of $c_T^2$, at least on the 
Friedmann-Lema\^{i}tre-Robertson-Walker (FLRW) 
background \cite{GeffGP}. 
However, the intrinsic vector mode is already present as the 
$F$ and $Y$ dependence in $G_2$, so we do not take them into 
account.

\subsection{Background equations and stability conditions}

We briefly review the equations of motion and stability conditions 
on the flat FLRW background described by the line element 
$ds^{2} = - dt^{2} + a^2(t) \delta_{ij}dx^{i}dx^{j}$, 
where $a(t)$ is the time-dependent scale factor. 
The vector field profile compatible with this background 
is given by $A^{\mu} = (\phi(t),0,0,0)$.
For the matter sector, we take into account nonrelativistic 
matter (density $\rho_m$ with vanishing pressure) and 
radiation (density $\rho_r$ and pressure $P_r=\rho_r/3$).
They obey the continuity equations 
$\dot{\rho}_m+3H \rho_m=0$ and 
$\dot{\rho}_r+4H \rho_r=0$, respectively, where a dot 
represents a derivative with respect to $t$ and 
$H=\dot{a}/a$ is the Hubble expansion rate.
The background equations of motion are \cite{cosmoGP}
\ba
& &
3M_{\rm pl}^2 H^2=\rho_{\rm DE}+\rho_m+\rho_r\,,
\label{back_eom_g00}\\
& &
M_{\rm pl}^2 \left( 2\dot{H}+3H^2 \right)
=-P_{\rm DE}-\frac{1}{3}\rho_r\,,
\label{back_eom_g11}\\
& &
\phi \left( G_{2,X}+3G_{3,X}H \phi \right)=0\,,
\label{backphi}
\ea
where $\rho_{\rm DE}$ and $P_{\rm DE}$ are the density and pressure of the ``dark'' component defined, respectively, by 
\be
\rho_{\rm DE}=-G_2\,,\qquad P_{\rm DE}=G_2-G_{3,X}\dot{\phi} \phi^2\,.
\ee
Here and in the following, we use the notation $G_{i,X} \equiv \partial G_i/\partial X$ with $i=2,3$.
The dark energy equation of state is defined as 
\be
w_{\rm DE} \equiv \frac{P_{\rm DE}}{\rho_{\rm DE}}
=-1+\frac{G_{3,X}\dot{\phi} \phi^2}{G_2}\,.
\label{wde}
\ee
The quantities $F$ and $Y$ vanish on the flat FLRW spacetime, so they do not contribute to the background equations of motion.

{}From Eq.~(\ref{backphi}), there is a branch of nonvanishing $\phi$ 
satisfying $G_{2,X}+3G_{3,X}H \phi=0$. In this case, the temporal vector 
component $\phi$, which is an auxiliary field, depends on $H$ alone. 
This gives rise to the existence of a de Sitter solution characterized by constant $\phi$ and $H$. 
If the vector field contributes to the late-time dark energy dynamics,
the temporal component $\phi$ increases toward the de Sitter solution 
($\dot{\phi}>0$ for the branch $\phi>0$).
{}From Eq.~(\ref{wde}), the cubic coupling $G_3(X)$ leads to the deviation 
of $w_{\rm DE}$ from $-1$. For the positive dark energy density 
$\rho_{\rm DE}=-G_2>0$ with $\dot{\phi}>0$ and $\phi>0$, 
$w_{\rm DE}$ is in the range $w_{\rm DE}<-1$ 
for the coupling $G_{3,X}>0$ 
before reaching the de Sitter solution.

In GP theories given by the action (\ref{action}), there are 
two polarized states of tensor perturbations, whose propagation speeds $c_T$ are both equivalent to 1  
on the flat FLRW background without ghosts \cite{cosmoGP}. 
In the small-scale limit, the conditions for the absence of 
ghosts and Laplacian instabilities of vector perturbations 
(characterized by two transverse modes)
are given, respectively, by \cite{GeffGP}
\ba
q_V &=& G_{2,F} + 2 G_{2,Y} \phi^{2}>0\,,
\label{qV}\\
c_V^2 &=& 1-\frac{2G_{2,Y}\phi^2}{q_V}>0\,.
\label{cV}
\ea
For scalar perturbations, there is a longitudinal scalar mode 
besides the perturbations arising from nonrelativistic 
matter and radiation. In the small-scale limit, 
the ghost and Laplacian instabilities 
of the longitudinal scalar are absent under the conditions
\ba
Q_{S} &=&
\frac{H^2 M_{\rm pl}^2}{(2H M_{\rm pl}^2-G_{3,X}\phi^3)^2}q_S>0\,,
\label{QS}\\
c_{S}^2 &=& \frac{2M_{\rm pl}^2}{q_S} \left[  
\left(2 G_{3,X} + G_{3,XX} \phi^{2} \right) \dot{\phi}
+G_{3,X}H \phi \right] 
+ \left( \frac{2M_{\rm pl}^2}{q_V\phi^{2}}
- 1 \right) \frac{G_{3,X}^{2} \phi^4}{q_{S}}>0\,,
\label{cscon}
\ea
where 
\be
q_{S}=
3G_{3,X} \phi^2 \left( G_{3,X}\phi^{2} 
-\frac{2M_{\rm pl}^2  \dot{H}}{\dot{\phi}} \right)\,.
\ee
To avoid the strong coupling problem, we also require that 
$Q_S$ does not approach 0 in any cosmological epoch.
For the matter sector, there are no Laplacian instabilities 
for $c_m^2>0$ and $c_r^2>0$, where $c_m$ and 
$c_r$ are the propagation speeds of nonrelativistic matter 
and radiation, respectively.

\subsection{Concrete dark energy model}

In this paper, we focus on the dark energy model 
proposed in Ref.~\cite{cosmoGP}, i.e., 
\be
G_{2} = b_{2} X^{p_{2}} + F\,,\qquad
G_{3} = b_{3} X^{p_{3}}\,,
\label{G23}
\ee
where $b_2, b_3, p_2, p_3$ are constants. 
The intrinsic vector mode is encoded as the Lagrangian $F$ 
in $G_2$, so we do not take into account the $Y$ dependence in $G_2$. 
In this case, we have 
\be
q_V=1\,,\qquad c_V^2=1\,,
\ee
and hence there are neither ghosts nor Laplacian instabilities 
of vector perturbations.
{}From Eq.~(\ref{backphi}), the nonvanishing temporal vector 
component $\phi$ obeys
\be
\phi^p H =-\frac{2^{p_{3}-p_{2}} b_{2}p_2}
{3 b_{3}p_{3}}={\rm constant}\,,
\label{phipre}
\ee
where 
\be
p \equiv 1-2p_2+2p_3\,.
\ee
In the following, we focus on the branch $\phi>0$,  
with the power $p$ satisfying
\be
p>0\,.
\label{pcon}
\ee
In this case, $\phi$ increases with the decrease of $H$.

The Hamiltonian constraint (\ref{back_eom_g00}) can be 
expressed in the form 
\be
\Omega_{m}= 1-\Omega_r-\Omega_{\rm DE}\,,
\ee
where 
\be
\Omega_{r} \equiv \frac{\rho_{r}}
{3 M_{\rm pl}^{2} H^2}\,,\qquad 
\Omega_{\rm DE} \equiv \frac{\rho_{\rm DE}}
{3 M_{\rm pl}^{2} H^2}=\frac{2^{-p_2}m^2u^{2p_2}}
{3H^2} \,.
\ee
Here we introduce
\be
b_2 \equiv -m^{2} M_{\rm pl}^{2 (1 - p_{2})}\,,\qquad 
u \equiv \frac{\phi}{M_{\rm pl}}\,,
\ee
with $m$ being a constant which has a dimension of mass. 
We require that $m^2>0$, i.e., $b_2<0$, to have a positive 
dark energy density.
On using Eqs.~(\ref{back_eom_g11}) and (\ref{phipre}), 
the density parameters $\Omega_{\rm DE}$ and 
$\Omega_{r}$ obey
\ba
\Omega_{\rm DE}' &=&
\frac{(1 + s)\, \Omega_{\rm DE}\, 
(3 + \Omega_{r} - 3 \Omega_{\rm DE })}
{1 + s\Omega_{\rm DE }} \,,
\label{auto_DE} \\
\Omega_{r}' &=& - \frac{\Omega_{r} [ 1 - \Omega_{r} + (3 + 4 s) \Omega_{\rm DE}]}{1 + s \Omega_{\rm DE}}\,,
\label{auto_r}
\ea
where a prime represents a derivative with respect to 
${\cal N}=\ln a$, and 
\be
s \equiv \frac{p_2}{p}=\frac{p_2}{1-2p_2+2p_3}\,.
\ee
To avoid the possible divergences of $\Omega_{\rm DE}$ and 
$\Omega_{r}$ arising from the denominators in 
Eqs.~(\ref{auto_DE}) and (\ref{auto_r}), we require 
the condition $s>-1$ for $0 < \Omega_{\rm DE} \le 1$.
The dark energy equation of state (\ref{wde}) 
is expressed as 
\be
w_{\rm DE} = -\frac{3 (1 + s) + s \Omega_{r}}{3 (1 + s \Omega_{\rm DE})}\,,
\label{wdef}
\ee
and hence $w_{\rm DE}=-1-s$ during the matter 
dominance ($\Omega_{\rm DE}=0$ and $\Omega_r=0$).
The ratio $s=p_2/p$, which quantifies the deviation of 
$w_{\rm DE}$ from $-1$, plays a key role in determining
the dark energy dynamics before reaching the de Sitter 
solution characterized by $\Omega_{\rm DE}=1$ and 
$\Omega_r=0$.

The no-ghost condition (\ref{QS}) of scalar perturbations yields
\be
Q_{S} = \frac{m^{2} p^{2} s}{(1 - ps \Omega_{\rm DE})^{2}}\,
\lambda^{\frac{2 (ps - 1)}{p (1 + s)}}
2^{-\frac{s (1 + p)}{1 + s}} 
3^{\frac{ps -1}{p (1 + s)}}
\left( 1 + s \Omega_{\rm DE} \right)\,
\left( \Omega_{\rm DE} \right)^{\frac{ps - 1}{p (1 + s)}} 
> 0\,,
\label{QS_model}
\ee
where 
\be
\lambda \equiv \left( \frac{\phi}{M_{\rm pl}} \right)^{p} \frac{H}{m}
=u^p \frac{H}{m}\,.
\label{lamdef}
\ee
{}From Eq.~(\ref{phipre}), the variable $\lambda$
is constant. The scalar ghost is absent for 
\be
s>0\,,
\label{scon}
\ee
under which the dark energy equation of state (\ref{wdef}) 
during the radiation and matter eras is in the range 
$w_{\rm DE}<-1$. In order to avoid the approach of 
$Q_S$ to 0 in the asymptotic past 
($\Omega_{\rm DE} \to 0$), we require that  
$(ps-1)/[p(1+s)] \leq 0$. Under the conditions 
(\ref{pcon}) and (\ref{scon}), this translates to
\be
0 < ps \leq 1\,.
\label{pscon1}
\ee
In other words, the power $p_{2}$ in $G_{2}$ 
is bounded as $0 < p_{2} \leq 1$.

From Eq.~(\ref{cscon}), the Laplacian instability of scalar 
perturbations is absent for
\be
c_{S}^{2} =  \frac{6 ps + 5 p - 3 + (2 ps + p - 1) \Omega_{r}
+ [3 - 3 p - 2 p s (2 + p)] \Omega_{\rm DE} - 2 p^{2} s^{2} \Omega_{\rm DE}^{2}}{6 p^{2} (1 + s \Omega_{\rm DE})^{2}}
+ \frac{2 s \Omega_{\rm DE}}{3 (1 + s \Omega_{\rm DE}) u^{2} q_V}
> 0\,,
\label{cS2_model}
\ee
where $q_V=1$ for the model (\ref{G23}). 
Since the normalized temporal vector component can 
be expressed as $u=(2^{p_2} 3\lambda^2 \Omega_{\rm DE})^{1/[2p(1+s)]}$, the last term 
of Eq.~(\ref{cS2_model}) is proportional to 
$(\Omega_{\rm DE})^{[p(1+s)-1]/[p(1+s)]}$.
To avoid the divergence of $c_S^2$ in the asymptotic past, 
we further impose the condition 
\be
p \left( 1+s \right) \geq 1\,.
\label{pscon2}
\ee
During the radiation, matter, and de Sitter eras, 
Eq.~(\ref{cS2_model}) reduces to 
\ba
& &
\left(c_{S}^{2}\right)_{\rm rad} \to \frac{4 ps  + 3 p - 2 }
{3 p^{2}} \,,\label{csra}\\
& &
\left(c_{S}^{2}\right)_{\rm mat} \to \frac{6 ps + 5 p - 3 }
{6 p^{2}} \,,\label{csma}\\
& &
\left(c_{S}^{2}\right)_{\rm dS} \to \frac{1}{3} 
\left[  \frac{1 - ps}{p (1+s)} +
\left( \frac{2}{3^{1/p} \lambda^{2/p}} \right)^{1/(1+s)}
\frac{s}{(1+s) q_{V}} \right] \,,\label{csds}
\ea
respectively. On using the conditions (\ref{qV}), (\ref{scon}), (\ref{pscon1}), and (\ref{pscon2}),
it follows that $(c_S^2)_{\rm rad}$, $(c_S^2)_{\rm mat}$, 
and $(c_S^2)_{\rm dS}$ are all positive.

\section{Cosmological perturbations and ISW-galaxy cross-correlations}
\label{sec3}

To confront the dark energy model in GP theories with the observations of 
RSDs and ISW-galaxy cross-correlations, we need to  study the evolution 
of matter density perturbations and gravitational potentials. 
For this purpose, we consider a perfect fluid of nonrelativistic matter with  
the sound speed squared $c_{m}^{2}$ close to $+0$. 
We introduce the matter perturbation $\delta \rho_m$ and the velocity potential $v$ 
in terms of the Schutz-Sorkin action \cite{Sorkin} along the lines of Ref.~\cite{GeffGP}.
The gauge-invariant matter density contrast is defined by 
\be
\delta \equiv \frac{\delta\rho_{m}}{\rho_{m}} + 3 H v\,.
\ee
For the gravity sector, we consider the linearly perturbed line element 
in the flat gauge given by 
\be
ds^2=-\left( 1+2\alpha \right) dt^2+2 \nabla_i B dt dx^i
+a^2(t) \delta_{ij}dx^i dx^j\,,
\ee
where $\alpha$ and $B$ are scalar metric perturbations. 
We also introduce the two Bardeen gravitational 
potentials \cite{Bardeen}:
\be
\Psi \equiv \alpha + \dot{B}\,, \qquad
\Phi \equiv H B \,. 
\ee
The gravitational potential associated with the bending of 
light rays is defined by 
\be
\psi_{\rm ISW} \equiv \Psi-\Phi\,,
\ee
which plays a key role for the ISW effect in CMB measurements.

In Fourier space with the comoving wave number $k=|{\bm k}|$, we relate $\Psi$ and $\Phi$ 
with the matter density contrast $\delta$, as 
\ba
\frac{k^2}{a^2} \Psi &=& - 4 \pi G \mu \rho_{m} \delta\,,
\label{Psieq} \\
\frac{k^{2}}{{a^{2}}} \psi_{\rm ISW} 
&=& - 8 \pi G \Sigma \rho_{m} \delta\,,
\label{psiISW}
\ea
where $G$ is the Newton gravitational constant. 
The quantities $\mu$ and $\Sigma$ are dimensionless (positive) 
gravitational couplings felt by matter and light, 
respectively \cite{Amen08,Song,Tate,Zhao,Bean,Sil2}.
We can express $\Sigma$ in the form 
\be
\Sigma=\frac{\mu}{2} \left( 1+\eta \right)\,,
\ee
where $\eta \equiv -\Phi/\Psi$ is the gravitational slip parameter.

\subsection{Cubic-order GP theories}

We first briefly review the gravitational couplings 
and the evolution of matter perturbations in GP theories. 
In Fourier space, the density contrast $\delta$ and 
the velocity potential $v$ obey
\ba
& &
\dot{\delta}-3\dot{\cal V}=-\frac{k^2}{a^2} 
\left( B+v \right)\,,\label{maeq1}\\
& &
\dot{v}=\alpha\,,\label{maeq2}
\ea
where ${\cal V} \equiv Hv$. 
Taking the time derivative of Eq.~(\ref{maeq1}) and using Eq.~(\ref{maeq2}), 
it follows that 
\be
\ddot{\delta} + 2 H \dot{\delta} + \frac{k^{2}}{a^{2}} \Psi 
= 3 \ddot{{\cal V}} + 6 H \dot{{\cal V}} \,.
\label{eq_delta}
\ee
For the theories in which the matter growth rate is not significantly 
different from that in GR, $\dot{\delta}$ is at most of order 
$H \delta$. Then, from Eq.~(\ref{maeq1}), the velocity potential can 
be estimated as
$|{\cal V}| \lesssim (aH/k)^2 |\delta|$. 
For the perturbations deep inside the Hubble radius ($k^2 \gg a^2 H^2$), 
the two terms on the right-hand side of Eq.~(\ref{eq_delta}) 
can be neglected relative to those on the left-hand side.
In this case, Eq.~(\ref{eq_delta}) reduces to 
\be
\ddot{\delta} + 2 H \dot{\delta} - 4 \pi G \mu \rho_{m} \delta 
\simeq 0\,,
\label{eqdelta_quasi}
\ee
where we used Eq.~(\ref{Psieq}).

The dimensionless gravitational coupling $\mu$ is known by 
solving the other perturbation equations of motion derived in Ref.~\cite{cosmoGP}. 
For the modes deep inside the sound horizon 
$(c_{S}^{2}k^{2}/a^{2} \gg H^{2})$, we can resort to the 
so-called quasistatic approximation under which the dominant terms 
in the perturbation equations are those containing 
$\delta \rho_m$ and $k^2/a^2$ \cite{Sta00,DKT11}.
Under the quasistatic approximation, the analytic expressions of $\mu$ and $\eta$ were 
already derived in the literature see [Eqs.~(5.29) and (5.30) of Ref.~\cite{GeffGP}]. 
In cubic-order GP theories given by the action (\ref{action}), we have 
\be
\mu=\Sigma=1 + \frac{(\phi^{2} G_{3,X})^{2}}{q_{S} c_{S}^{2}}\,,\qquad 
\eta=1\,,
\label{Geff}
\ee
and hence there is no gravitational slip. 
In this case, the gravitational interactions felt by matter and light are equivalent to 
each other. Under the absence of ghosts and Laplacian instabilities of 
scalar perturbations, the gravitational interactions are enhanced 
($\mu=\Sigma>1$) compared to those in GR ($\mu=\Sigma=1$). 
Since $\mu$ and $\Sigma$ do not depend on $k$ under the quasistatic 
approximation, the matter density contrast $\delta$ evolves 
in a scale-independent way according to Eq.~(\ref{eqdelta_quasi}) 
for the perturbations inside the sound horizon.

\subsection{ISW-galaxy cross-correlations}
\label{ISWsub}

In this section, we derive the power spectrum of ISW-galaxy cross-correlations 
in a general way without specifying gravitational theories. 
During the matter era the gravitational potential $\psi_{\rm ISW}$ does not 
typically change in time,  
but the dominance of dark energy leads to the variation of  $\psi_{\rm ISW}$ 
at low redshifts. This leaves an imprint on temperature anisotropies of CMB photons
freely streaming from the last scattering surface to today. 
The ISW contribution $\Delta T_{\rm ISW}$ to the CMB temperature perturbation 
divided by the average temperature $T$ can be quantified by 
the integral with respect to the redshift $z=1/a-1$, such that 
\be
\frac{\Delta T_{\rm ISW}(\hat{n})}{T} 
= - \int_{0}^{z_{r}} dz \frac{\partial \psi_{\rm ISW}}{\partial z}\,,
\ee
where $\hat{n}$ is a unit vector along the line of sight and 
$z_{r}$ is the redshift at recombination.

The clustering of galaxies occurs by the growth of matter density contrast $\delta$. 
For the theories in which the dimensionless 
gravitational coupling $\mu$ does not depend on the wave number ${\bm k}$,  
we can express the Fourier-space perturbation $\delta$ 
at the redshift $z$ in the form 
\be
\delta(z, \bm{k})=\frac{D(z)}{D_{0}} \delta(0, \bm{k})\,,
\label{delD}
\ee
where we introduce the growth factor $D(z)$ 
with today's value $D_{0}\equiv D(z=0)$. 
The fluctuations in the angular distribution of galaxies can be 
quantified as
\be
\frac{\Delta N_{\rm Galaxy}(\hat{n})}{N} = \int_{0}^{z_{r}} 
dz\, b_{s}^{A}\, \phi^{A}(z)\, \delta(z,  \hat{n}\chi(z))\,,
\ee
where $b_{s}^{A}$ is a bias factor, 
$\phi^{A}(z)$ is a window function, and 
$\chi=\int_0^z H^{-1}(\tilde{z})d\tilde{z}$ is a comoving distance.
The label $A$ stands for different galaxy catalogues.
For the window function, we choose the following form \cite{ISWdata}:
\be
\phi^{A}(z) = \frac{\beta}{\Gamma[(\alpha+1)/\beta]} \left( \frac{z}{z_{0}} \right)^{\alpha} 
\exp\left[ -\left( \frac{z}{z_{0}} \right)^{\beta} \right]\,,
\label{window}
\ee
where $\Gamma[x]$ is the gamma function and 
$\alpha, \beta, z_{0}$ are positive constants.
The values of these constants are different depending on the 
galaxy surveys.
The function (\ref{window}), which is positive, satisfies the normalization 
$\int_{0}^{\infty} dz\, \phi^{A}(z) = 1$, and it has a peak around $z=z_0$. To confront our model with the observational 
data, we select the two galaxy surveys: the 2 Micron All-Sky Survey (2MASS) and SDSS, in which case the window functions for 
galaxy bins are considerably peaked at particular 
redshifts \cite{ISWdata}.
The 2MASS galaxy catalogue can be fitted by the window 
function (\ref{window}) with 
$(z_{0}, \alpha, \beta) = (0.072, 1.901, 1.752)$.
For the SDSS catalogue, we choose the parameters 
$(z_{0}, \alpha, \beta) = (0.113, 3.457, 1.197)$.

In the following, we also assume that the bias $b_s^A$ is 
scale independent as well as time independent in the range 
of redshift intervals allowed by $\phi^{A}(z)$.
This is a reasonable assumption for galaxy catalogues 
with the peaked window 
function mentioned above.

Let us consider a perturbation $X(z,\chi\hat{n})$ that depends on $z$ and 
the product of comoving distance $\chi$ and unit vector 
$\hat{n}$. Then, the perturbation $X(\hat{n})$, 
which corresponds to the integration of $X(z,\chi\hat{n})$ with 
respect to $z$ from $z=0$ to $z=\infty$, 
can be expanded in terms of spherical harmonics 
$Y_{lm} (\hat{n})$, as 
\be
X(\hat{n})=\int_{0}^{\infty} dz\,
X(z,\chi \hat{n})=\sum_{l,m} a_{lm}^X
Y_{lm} (\hat{n})\,,
\ee
where $a_{lm}^X=\int d\Omega X(\hat{n}) Y_{lm}^{*} (\hat{n})$ with the solid angle $\Omega$. 
The Fourier-series expansion of $X(z,\chi\hat{n})$ is given by 
\be
X(z,\chi\hat{n})=\int \frac{d^3k}{(2\pi)^3}
X(z,{\bm k}) e^{i{\bm k}\cdot \chi \hat{n}}\,.
\ee
On using the relation 
$\int d\Omega e^{i{\bm k}\cdot {\bm r}}Y_{lm}^*(\hat{r})
=4\pi i^l j_l (kr)Y_{lm}^*(\hat{k})$, where 
$\hat{r}={\bm r}/r$, $\hat{k}={\bm k}/k$, with 
the spherical Bessel function $j_l(x)$, the coefficient $a_{lm}^X$ 
is expressed as
\be
a_{lm}^X=\frac{i^l}{2\pi^2} \int dz \int d^3k\,
X(z,{\bm k}) j_l (k \chi)Y_{lm}^*(\hat{k})\,.
\label{alm}
\ee
The coefficients $a_{lm}^{\rm ISW}$ and $a_{lm}^{\rm Galaxy}$, which 
are associated with the ISW signal and galaxy 
clusterings, respectively,  
can be derived by substituting 
$X \to -\partial \psi_{\rm ISW}/\partial z$ and 
$X \to b_s^A \phi^A (z) (D(z)/D_0)\delta (0,{\bm k})$ 
into Eq.~(\ref{alm}).
In doing so, we exploit the properties 
$1+z=e^{-{\cal N}}$ and $dz/d{\cal N}=-e^{-{\cal N}}$ between 
the redshift $z$ and the $e$-folding number ${\cal N}=\ln a$.
Then, it follows that \cite{MTMG_ISW}
\ba
a_{lm}^{\rm ISW}
&=&-\frac{i^l}{2\pi^2 D_0} \int d{\cal N}_1 
\int d^3 k_1\,Z_{\rm ISW} ({\cal N}_1) \delta (0,{\bm k}_1) 
j_l (k_1 \chi_1) Y_{lm}^* (\hat{k}_1)\,,\label{alm1}\\
a_{lm}^{\rm Galaxy}
&=& -\frac{i^l}{2\pi^2 D_0} \int d{\cal N}_2 e^{-{\cal N}_2}
\int d^3 k_2\,b_s^A \phi^A ({\cal N}_2)
D({\cal N}_2)\delta (0,{\bm k}_2) 
j_l (k_2 \chi_2) Y_{lm}^* (\hat{k}_2)\,,\label{alm2}
\ea
where $Z_{\rm ISW}$ is defined by 
\be
\frac{\partial \psi_{\rm ISW}}{\partial {\cal N}}=
Z_{\rm ISW} ({\cal N}, k) \frac{\delta (0,{\bm k})}{D_0}\,.
\label{Zdef}
\ee

The cross-correlation between the ISW signal in CMB and 
the galaxy fluctuations is quantified as
\be
\left< \frac{\Delta T_{\rm ISW}(\hat{n}_1)}{T} 
\frac{\Delta N_{\rm Galaxy}(\hat{n}_2)}{N}\right>
= \sum_{l = 0}^{\infty} \frac{2 l + 1}{4 \pi} 
C_{l}^{\rm IG} \mathcal{P}_{l}(\cos\theta) \,,
\ee
where $\mathcal{P}_{l}$ is the Legendre polynomial 
with the angle $\theta$ between the unit vectors 
$\hat{n}_1$ and $\hat{n}_2$, 
and $C_{l}^{\rm IG}$ is the ISW-galaxy cross-correlation amplitude 
given by 
\be
C_{l}^{\rm IG} = \left< a_{lm}^{\rm ISW} 
( a_{lm}^{\rm Galaxy})^{*} \right>\,.
\label{Cdef}
\ee
Substituting Eqs.~(\ref{alm1}) and (\ref{alm2}) into Eq.~(\ref{Cdef}), 
we obtain
\be
C_{l}^{\rm IG}
=\frac{2b_{s}^{A} }{\pi D_{0}^{2}}
\int_{k_m}^{k_M} dk \, k^{2} \, P_{\delta} (k)
\int_{\mathcal{N}_{i}}^{0} 
d\mathcal{N}_{1} Z_{\rm ISW}(\mathcal{N}_1, k)   
j_{l}[k\chi(\mathcal{N}_{1})] 
\int_{\mathcal{N}_{i}}^{0}  
d\mathcal{N}_{2} e^{-\mathcal{N}_{2}} \phi^{A}(\mathcal{N}_{2}) 
D(\mathcal{N}_{2})   j_{l}[k\chi(\mathcal{N}_{2})] \,,
\label{Cl}
\ee
where $k_m$ and $k_M$ are minimum and maximum wave numbers, respectively, 
${\cal N}_i$ is the initial value of ${\cal N}$ in the deep matter era, 
and $P_{\delta}$ is the matter power spectrum defined by 
\be
\left< \delta(0, \bm{k}_1) \delta^*(0, \bm{k}_2) \right>
= (2 \pi)^{3} \delta_{D}^{(3)}(\bm{k}_1- \bm{k}_2) 
P_{\delta}(k_1) \,.
\ee
Similarly, the galaxy-galaxy correlation amplitude can be computed as
\ba
\hspace{-0.9cm}
C_l^{\rm GG}
&=& \left< a_{lm}^{\rm Galaxy} (a_{lm}^{\rm Galaxy})^* \right> 
\nonumber \\
\hspace{-0.9cm}
&=& \frac{2(b_{s}^{A})^2}{\pi D_{0}^{2}}
\int_{k_m}^{k_M} dk \, k^{2} \, P_{\delta} (k)
\left( \int_{\mathcal{N}_{i}}^{0}  
d\mathcal{N} 
e^{-\mathcal{N}}\phi^{A}(\mathcal{N}) 
D(\mathcal{N})   j_{l}[k\chi(\mathcal{N})] 
\right)^2\,.
\label{ClGG}
\ea
On using the transfer function $T_m(k)$ from the deep radiation era to the matter-dominated epoch, the matter power spectrum can be expressed as
\be
P_{\delta}(k)=2\pi^2 \delta_H^2 T_m^2(k) 
\left( \frac{k}{H_0} \right)^{n_s} H_0^{-3}\,,
\label{Pdelta}
\ee
where $\delta_{H}$ and $n_s$ are the amplitude and the spectral 
index of primordial scalar perturbations, respectively. 
We employ the transfer function $T_m(k)$ advocated by 
Eisenstein and Hu \cite{EHu1,EHu2}.
Substituting Eq.~(\ref{Pdelta}) into Eqs.~(\ref{Cl}) and 
(\ref{ClGG}), it follows that  
\ba
\hspace{-1cm}
C_{l}^{\rm IG} &=& 
4\pi b_{s}^{A}\,\bar{\delta}_H^2 
\int_{k_m}^{k_M} \frac{dk}{k}
\left( \frac{k}{H_0} \right)^{n_s+3} 
T_m^2(k) \int_{\mathcal{N}_{i}}^{0} 
d\mathcal{N}_{1} Z_{\rm ISW}(\mathcal{N}_1, k)   
j_{l}[k\chi(\mathcal{N}_{1})] 
\int_{\mathcal{N}_{i}}^{0}  
d\mathcal{N}_{2} e^{-\mathcal{N}_{2}}  \phi^{A}(\mathcal{N}_{2}) 
D(\mathcal{N}_{2})   j_{l}[k\chi(\mathcal{N}_{2})],
\label{Cl2}\\
\hspace{-1cm}
C_{l}^{\rm GG} &=& 
4\pi (b_{s}^{A})^2\,\bar{\delta}_H^2 
\int_{k_m}^{k_M} \frac{dk}{k}
\left( \frac{k}{H_0} \right)^{n_s+3}
T_m^2(k) 
\left( \int_{\mathcal{N}_{i}}^{0}  
d\mathcal{N} 
e^{-\mathcal{N}}\phi^{A}(\mathcal{N}) 
D(\mathcal{N})   j_{l}[k\chi(\mathcal{N})] 
\right)^2\,,
\label{ClGG2}
\ea
where 
\be
\bar{\delta}_H \equiv \frac{\delta_H}{D_0}\,.
\ee

The quantity $Z_{\rm ISW}$ plays a key role for determining the sign 
of $C_l^{\rm IG}$. We recall that the gravitational potential 
$\psi_{\rm ISW}$ is related to $\delta$ according to Eq.~(\ref{psiISW}). 
The density $\rho_m$ is given by 
$\rho_m=3M_{\rm pl}^2 H_0^2 \Omega_{m0} (1+z)^3$, where 
$\Omega_{m0}$ is today's density parameter of nonrelativistic matter. 
Using the relation (\ref{delD}), we can express $\psi_{\rm ISW}$ 
in the form 
\be
\psi_{\rm ISW} = - \frac{3 H_{0}^{2} \Omega_{m0}}{k^{2}}\, 
e^{-\mathcal{N}} D \Sigma \frac{\delta(0,\bm{k})}{D_{0}} \,.
\label{psiISW2}
\ee
Taking the $\mathcal{N}$-derivative of Eq.~(\ref{psiISW2}) and 
comparing it with Eq.~(\ref{Zdef}), it follows that 
\be
Z_{\rm ISW}(\mathcal{N}, k) =
\frac{3 H_{0}^{2} \Omega_{m0}}{k^{2}}\, e^{-\mathcal{N}}
D \Sigma  \mathcal{F} \,,
\label{ZISW}
\ee
where we introduced the following quantity: 
\be
\mathcal{F} \equiv 1 - \frac{D'}{D} - \frac{\Sigma'}{\Sigma}
=1-\left( \ln D \Sigma \right)'\,.
\label{Fdef}
\ee
Substituting Eq.~(\ref{ZISW}) into Eq.~(\ref{Cl2}), 
we obtain 
\be
C_l^{\rm IG}
=  \frac{12 \pi b_{s}^{A} 
\bar{\delta}_{H}^{2} \Omega_{m0}}{H_0^{n_s+1}} 
\int dk\,k^{n_s} T_m^2(k) \int_{{\cal N}_i}^0 d{\cal N}_1 e^{-{\cal N}_1} 
D({\cal N}_1) \Sigma({\cal N}_1) {\cal F}({\cal N}_1)
j_{l}(k\chi_1) 
\int_{{\cal N}_i}^0 d{\cal N}_2 e^{-{\cal N}_2}
D({\cal N}_2) \phi^{A}(\mathcal{N}_{2})  
j_{l}(k\chi_2),
\label{ClIG}
\ee 
where $\chi_i \equiv \chi(\mathcal{N}_{i})$ 
with $i=1,2$. 

For the large wave number $k$, it is useful to employ the
following Limber approximation for an arbitrary 
$k$-dependent function $f(k)$:
\be
\int dk\,k^{2} f(k) j_{l}(k\chi_{1})j_{l}(k\chi_{2}) 
\simeq \frac{\pi}{2} \frac{\delta(\chi_{1}-\chi_{2})}{\chi_{1}^{2}}
f\left(\frac{l_{12}}{\chi_{1}}\right) \,,
\label{Limber}
\ee
where $l_{12} \equiv l + 1/2$. 
Applying the approximation (\ref{Limber}) to Eq.~(\ref{ClIG}) and 
using Eq.~(\ref{Pdelta}) and the relation $d{\cal N}/d\chi=-aH$, we obtain 
\be
C_{l}^{\rm IG} \simeq \frac{6 \pi^{2} b_{s}^{A} 
\bar{\delta}_{H}^{2} \Omega_{m0}}{l_{12}^{2}}
\int_{\mathcal{N}_{i}}^{0} d\mathcal{N} e^{-\mathcal{N}} \frac{H}{H_{0}}
\left( \frac{l_{12}}{\bar{\chi}} \right)^{n_{s}} 
T_{m}^2 \left(\frac{l_{12}H_{0}}{\bar{\chi}}\right)
\phi^{A} D^2 \Sigma \mathcal{F} \,,
 \label{C_IG_limber}
\ee
where $\bar{\chi} \equiv H_0 \chi$.

The negative ISW-galaxy cross-correlation ($C_l^{\rm IG}<0$) 
can occur for the models in which ${\cal F}<0$ at low redshifts, 
which translates to 
\be
\left( \ln D \Sigma  \right)'>1\,.
\label{SigD}
\ee
Since $C_l^{\rm IG}$ is the integral with respect to ${\cal N}$ from 
the deep matter era to today, the condition (\ref{SigD}) is necessary 
but not sufficient for realizing $C_l^{\rm IG}<0$. 
As we will see in Sec.~\ref{sec3C}, even if ${\cal F}$ becomes negative 
at low redshifts, there are cases in which $C_l^{\rm IG}$ is positive.
 
Writing the factor $D'/D$ in Eq.~(\ref{Fdef}) in terms of the matter 
density parameter $\Omega_m$ and the growth index $\gamma$, 
as $D'/D=(\Omega_m)^{\gamma}$, it follows that 
${\cal F}=1-(\Omega_m)^{\gamma}-\Sigma'/\Sigma$.
In the $\Lambda$CDM model, the growth index is well 
approximated by $\gamma \simeq 0.55$ 
at low redshifts \cite{Wang}.
Since $\Sigma=1$ in this case, we have 
${\cal F}=1-(\Omega_m)^{\gamma}>0$ and hence 
the ISW-galaxy cross-correlation is positive
in the $\Lambda$CDM model.

In modified gravity theories the growth index is generally 
different from $0.55$. 
In $f(R)$ gravity, for example, it is in the range 
$0.40 \lesssim \gamma \lesssim 0.55$ \cite{Moraes}. 
The observational data of RSDs and the clustering of 
luminous red galaxies placed the bound $\gamma=0.56 \pm 0.05$ 
for constant $\gamma$ \cite{Pouri}, so the quantity 
$1-(\Omega_m)^\gamma$
is positive for the redshift $z$ relevant to the galaxy surveys 
($z \lesssim 2$). To realize the negative ISW-galaxy 
cross-correlation, it is at least necessary to satisfy the condition 
\be
\Sigma'>0
\ee
at low redshifts.

Before closing this subsection, we explain how to compute 
the quantities $\bar{\delta}_H$ and $b_s^A$ in the expression 
of Eq.~(\ref{C_IG_limber}).
The $k$-integrals in Eqs.~(\ref{Cl2}) and (\ref{ClGG2}) contain terms 
that depend on the window function. To extract such contributions,
we introduce the following quantity: 
\be
I \equiv \int_{k_m}^{k_M} \frac{dk}{k} 
\left( \frac{k}{H_0} \right)^{n_s+3}
\left[ T_m(k)\,w_{\rm TH} (8h^{-1}, k) \right]^2 \,,
\label{Idef}
\ee
where $w_{\rm TH}$ is the top-hat function defined by 
\be
w_{\rm TH}(r,k)=\frac{3[\sin (kr)-kr \cos (kr)]}
{(kr)^3}\,.
\ee
The quantity (\ref{Idef}) is evaluated at the scale 
$r=8h^{-1}$ Mpc, where $h$ is the normalized Hubble constant 
given by $H_{0} = 100\, h\, {\rm km\, s^{-1}\, Mpc^{-1}}$.
For the scalar spectral index $n_s$, we choose the 
best-fit value $n_s= 0.9649$ constrained from 
the Planck 2018 data \cite{Planck2018}.
 
We define today's amplitude of overdensity at the scale 
$8h^{-1}$ Mpc, as
\be
\sigma_8 (0) \equiv \delta_H \sqrt{I}\,.
\ee
{}From Eq.~(\ref{delD}) the value of $\sigma_8$
at the initial redshift $z_i$ in the deep matter era 
is related to $\sigma_8(0)$, as 
$\sigma_8(z_i)=\sigma_8(0)\,D(z_i)/D_0$. 
Then, the perturbation $\bar{\delta}_H=\delta_H/D_0$ 
is expressed as 
\be
\bar{\delta}_H=\frac{\sigma_8 (z_i)}{D(z_i)} 
\frac{1}{\sqrt{I}}\,.
\label{delHre}
\ee
Provided that the evolution of perturbations in the deep matter 
era is close to that in the $\Lambda$CDM model, 
the initial growth factor can be chosen as 
$D(z_i)=a_i=e^{{\cal N}_i}$.
Today's growth factor $D_0$ is known by solving Eq.~(\ref{eqdelta_quasi}) for $\delta$. 
Since the scalar-field contribution to the dynamics of perturbations tends to 
be negligible at higher redshifts in our model,  
we choose the same early-time initial conditions as those in the 
$\Lambda$CDM model. In particular, we consider initial conditions 
for $\sigma_8(z_i)$, such that $\sigma_8(z_i)=\sigma_8(z_i)^{\Lambda {\rm CDM}}$, and find $\sigma_8(z_i)^{\Lambda {\rm CDM}}$ by using $\sigma_8(z_i)^{\Lambda {\rm CDM}}=\sigma_8(0)^{\Lambda {\rm CDM}}\,e^{\mathcal{N}_i}/D_0^{\Lambda {\rm CDM}}$. 
For $\sigma_{8}(0)^{\Lambda {\rm CDM}}$, we choose the Planck best-fit value $\sigma_{8}(0)^{\Lambda {\rm CDM}} = 0.811$ \cite{Planck2018}.
Since the initial condition for $\sigma_8 (z_i)$ is now fixed, 
the value of $\bar{\delta}_H$ is known from Eq.~(\ref{delHre}).

For the bias factor $b_s^A$, we normalize it by using the observed best-fit galaxy-galaxy correlation  
spectrum $C_l^{\rm GG}$.
The analysis of Ref.~\cite{Rassat} using the galaxy spectrum 
data of 2MASS surveys combined with the WMAP data showed that the best-fit value of bias is $b_s^{\rm 2MASS}=1.4$.
For the SDSS survey,  the galaxy spectrum is consistent 
with the WMAP best-fit $\Lambda$CDM 
cosmology with the bias factor
$b_s^{\rm SDSS}=1$ \cite{ISWdata}. 
Then, for each galaxy survey, we can compute the 
galaxy power spectrum $C^{\rm GG}_{l,{\rm best}}$ 
by using the best-fit bias and best-fit cosmological parameters constrained from WMAP. 
We write the power spectrum (\ref{ClGG2}) 
in the form $C_l^{\rm GG}=4\pi (b_s^A)^2 
\bar{\delta}_H^2 Y_l^{{\rm GG}, A}$ and define
the $\chi^2$ estimator:
\be
\chi_{{\rm bias},A}^2 \equiv 
\sum_{l=2}^{150} \left[ C^{\rm GG}_{l,{\rm best}}
-4\pi (b_s^A)^2 
\bar{\delta}_H^2 Y_l^{{\rm GG}, A} \right]^2\,.
\ee
The bias can be fixed by minimizing $\chi_{{\rm bias},A}^2$.
Solving $\partial \chi^2_{{\rm bias},A}/\partial  b_s^A=0$ for $b_s^A$, it follows that 
\be
b_s^A=\sqrt{\frac{\sum_{l}C^{\rm GG}_{l,{\rm best}}
Y_l^{{\rm GG}, A}}{4\pi \bar{\delta}_H^2
\sum_{l}(Y_l^{{\rm GG}, A})^2}}\,.
\label{bses}
\ee
Computing $b_s^A$ from Eq.~(\ref{bses}) 
for dark energy models in GP 
theories, we have confirmed that the bias depends only 
mildly on the model parameters (typically within a 
few percent difference). This means that, as in the 
minimal theory of massive gravity \cite{MTMG_ISW}, 
using the power spectrum $C^{\rm GG}_{l,{\rm best}}$ 
derived for the best-fit $\Lambda$CDM cosmology 
is a reasonable prescription for the bias estimation.

\subsection{ISW-galaxy cross-correlations in GP theories}
\label{sec3C}

Let us consider the dark energy model in GP theories 
characterized by the functions (\ref{G23}).
{}From Eq.~(\ref{Geff}), the quantities $\mu$ and 
$\Sigma$ are expressed as 
\be
\mu = \Sigma=
1 + \frac{ s\, \Omega_{\rm DE}}
{3(1 + s\, \Omega_{\rm DE}) c_{S}^{2}}\,.
\label{mu_model}
\ee
During the radiation and matter eras, the scalar 
propagation speed squares are given, respectively, by 
Eqs.~(\ref{csra}) and (\ref{csma}).
Since $\Omega_{\rm DE} \ll 1$ in these epochs, 
$\mu$ and $\Sigma$ are close to 1.

On using Eq.~(\ref{csds}) at the de Sitter solution 
($\Omega_{\rm DE}=1$), it follows that 
\be
\mu_{\rm dS} = \Sigma_{\rm dS}=
1 + \left[\frac{1 - ps}{ps} 
+\left( \frac{2}{3^{1/p}} \right)^{1/(1+s)} 
\frac{1}{\lambda_V}
\right]^{-1} \,,
\label{muds}
\ee
where 
\be
\lambda_V \equiv \lambda^{2/[p(1+s)]}
q_{V}= \lambda^{2/[p(1+s)]}\,.
\label{lambdaV}
\ee
In the last equality, we used the fact that the model 
(\ref{G23}) satisfies $q_V=1$ (under which there is no 
issue of the strong coupling problem).
The intrinsic vector mode affects $\mu_{\rm dS}$ and 
$\Sigma_{\rm dS}$ through the quantity 
$\lambda_V=\lambda^{2/[p(1+s)]}$. 
Since $\lambda>0$ and $p(1+s) \geq 1$, both $\mu_{\rm dS}$ 
and $\Sigma_{\rm dS}$ are larger than 1. 
In the limit $\lambda_V \to \infty$,  
Eq.~(\ref{muds}) reduces to 
$\mu_{\rm dS}=\Sigma_{\rm dS} \to 
1/(1-ps)$, which corresponds to the values in 
cubic-order Horndeski (scalar-tensor) theories.
In another limit $\lambda_V \to 0$, we have 
$\mu_{\rm dS}=\Sigma_{\rm dS} \to 1$ and hence 
the evolution of perturbations is similar to
that in GR. 

\begin{figure}
\begin{center}
\includegraphics[height=3.2in,width=3.5in]{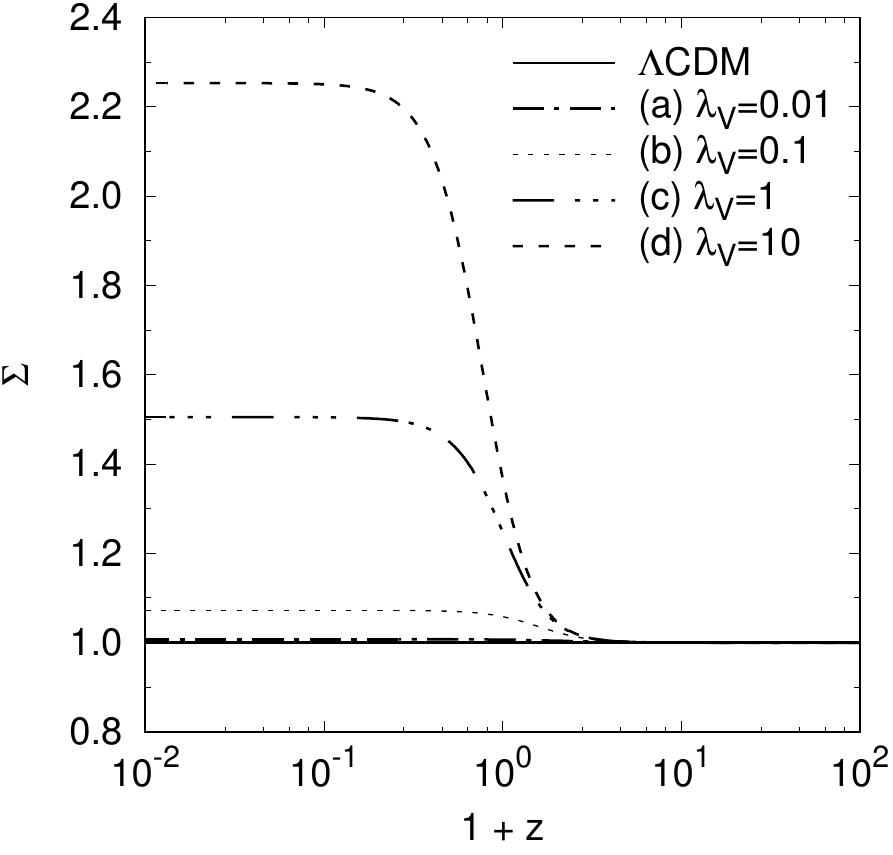}
\includegraphics[height=3.2in,width=3.5in]{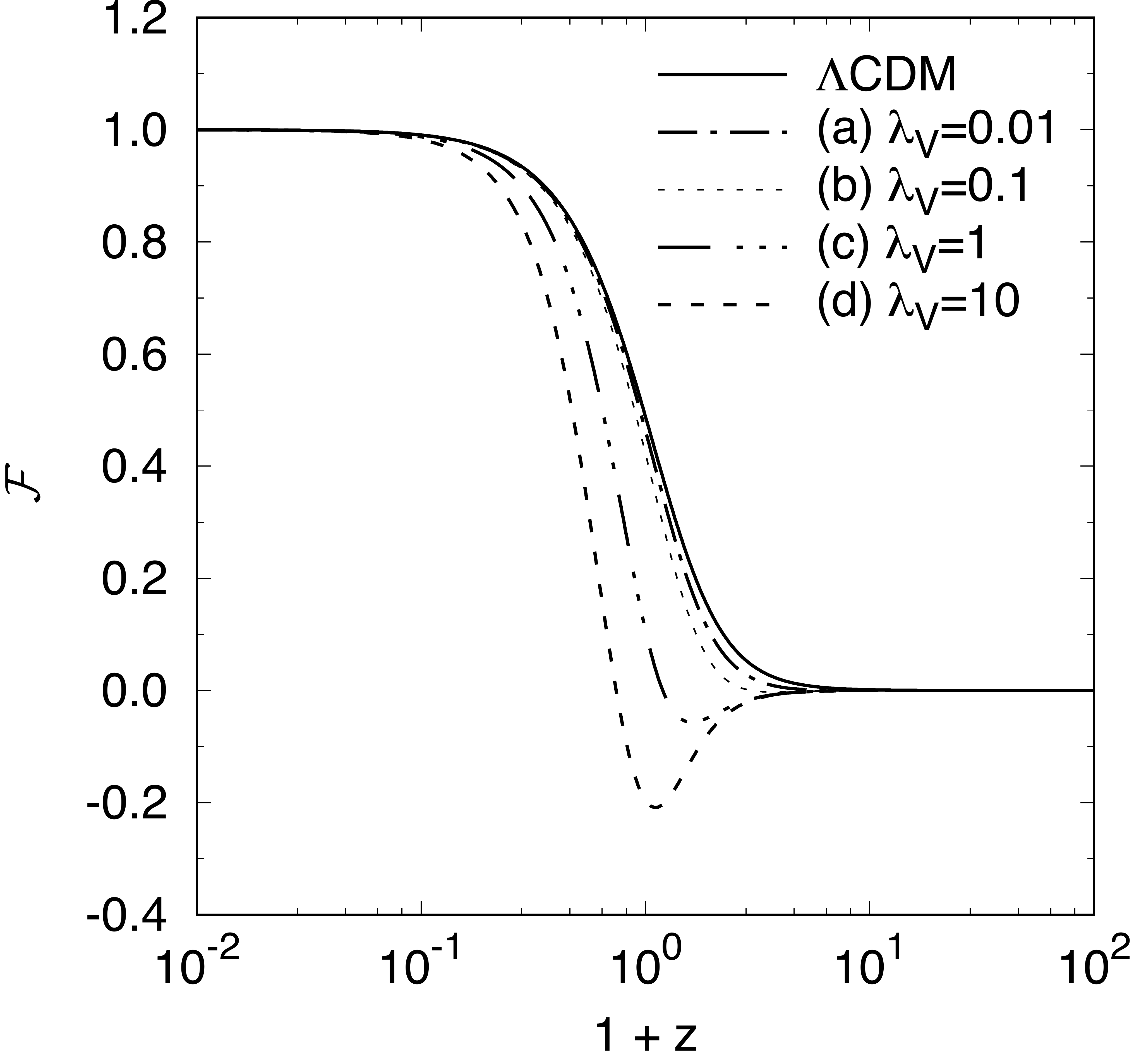}
\end{center}
\caption{\label{fig1}
Evolution of $\Sigma$ (left) and 
${\cal F}$ (right) versus $1 + z$ for $s=0.2$, 
$p=3$, and $\Omega_{m0}=0.32$ 
with four different values 
of $\lambda_V$: (a) $\lambda_V=0.01$, 
(b) $\lambda_V=0.1$, 
(c) $\lambda_V=1$, and (d) $\lambda_V=10$. 
The solid line corresponds to the evolution of $\Sigma$ 
and ${\cal F}$ in the $\Lambda$CDM model. 
For $\lambda_V \gtrsim 1$, the perturbation enters 
the region ${\cal F}<0$ at low redshifts. 
}
\end{figure}

\begin{figure}[htbp]
\begin{center}
\includegraphics[height=3.2in,width=3.4in]{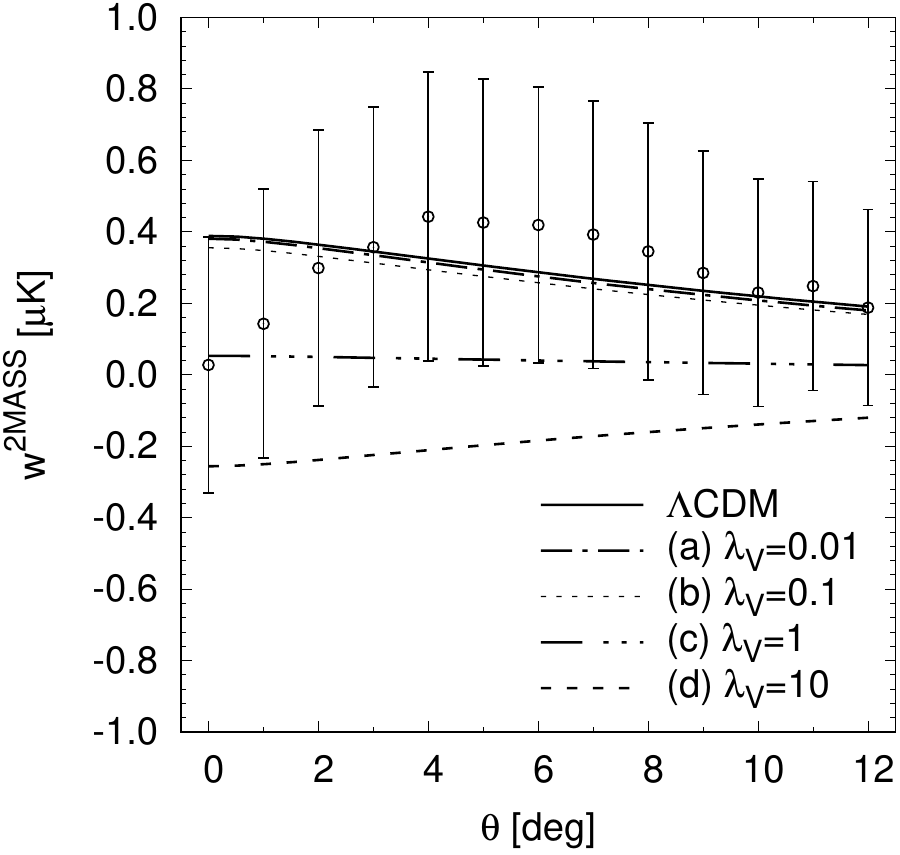}
\end{center}
\caption{\label{fig2}
The ISW-galaxy cross-correlation observable 
$w^{\rm 2MASS}$ versus 
the angle $\theta$ (representing the  deviation from 
the center of galaxy data sets) 
for the same model parameters as 
those used in Fig.~\ref{fig1} 
with $n_s=0.9649$, $\sigma_8(0)=0.811$, and $h=0.696$. 
We also show the data points of 2MASS measurements 
with error bars \cite{ISWdata} 
(derived by the jackknife error estimation method). 
}
\end{figure}

In the left panel of Fig.~\ref{fig1}, we show the evolution of 
$\Sigma~(=\mu)$ for four different values of $\lambda_V$
with $q_V=1$. The other model parameters are chosen to be 
$s=0.2$ and $p=3$ with today's matter density parameter 
$\Omega_m(z=0)=0.32$. 
In the $\Lambda$CDM model, the quantity $\Sigma$ 
is equivalent to 1 throughout the cosmological evolution. 
This case can be regarded as the limit $\lambda_V \to 0$ 
in Eq.~(\ref{muds}). 
As estimated from Eq.~(\ref{mu_model}), $\Sigma$ 
is close to 1 in the deep matter era for any value 
of $\lambda_V$ under consideration. 
The deviation of $\Sigma$ from 1 starts to occur at low redshifts. 
For larger $\lambda_V$, the deviation of $\Sigma$ 
from 1 tends to be more significant. 
This reflects the fact that, for 
increasing $\lambda_V$, the de Sitter value 
$\Sigma_{\rm dS}$ 
in Eq.~(\ref{muds}) gets larger, e.g., 
$\Sigma_{\rm dS}=1.07$ for $\lambda_V=0.1$ and 
$\Sigma_{\rm dS}=2.25$ for $\lambda_V=10$ in 
the numerical simulation of Fig.~\ref{fig1}.

In the right panel of Fig.~\ref{fig1}, we also plot the evolution of 
the quantity ${\cal F}$ defined by Eq.~(\ref{Fdef}) 
for the same model parameters as those used in 
the left panel. In all the cases the quantity ${\cal F}$ starts to evolve from 
the value close to $+0$ and finally approaches the asymptotic value 1, 
but the intermediate evolution of ${\cal F}$ is different depending 
on the parameter $\lambda_V$. 
In the $\Lambda$CDM model we have ${\cal F}>0$ throughout 
the cosmological evolution, so the ISW-galaxy cross-correlation is positive. 
In GP theories, the growth of $\Sigma$ occurs at low redshifts, in which case 
${\cal F}$ can be negative. 
With the model parameters used in Fig.~\ref{fig1}, the 
perturbation temporally enters the region ${\cal F}<0$ 
for $\lambda_V \gtrsim 1$. 
When $\lambda_V \gg 1$ the minimum value of ${\cal F}$
is largely negative, so it is expected that the strong negative ISW-galaxy 
cross-correlation occurs.

The observable associated with the ISW-galaxy 
cross-correlation is given by 
\be
w^A(\theta) \equiv T_{\rm CMB} \sum_{l = 0}^{\infty} 
\frac{2 l + 1}{4 \pi} C_{l}^{{\rm IG},A} \mathcal{P}_{l}(\cos \theta)\,,
\label{w_def}
\ee
where $T_{\rm CMB} = 2.7255\,{\rm K}$ and $\theta$ is 
the angle characterizing the deviation from the center of galaxy data sets.
For the calculation of $C_{l}^{{\rm IG}}$, we employ the formula 
(\ref{C_IG_limber}) derived under the Limber approximation.
In Fig.~\ref{fig2}, we plot $w^{\rm 2MASS}$ versus $\theta$ 
corresponding to 2MASS 
galaxy surveys for the same model 
parameters as those adopted in Fig.~\ref{fig1} with $n_s=0.9649$, 
$\sigma_8(0)=0.811$, and $h=0.696$. 
In Fig.~\ref{fig2}, the data points from the 
2MASS survey  are also shown with error bars. 
We note that the bias factor has been computed 
according to the formula (\ref{bses}) for the 
window function (\ref{window}) 
fitted to the 2MASS survey. 
The numerical values of $b_s^{\rm 2MASS}$ for 
$\lambda_{V}=0.01, 0.1, 1, 10$
are $1.497$, $1.487$, $1.475$, $1.471$, respectively, 
so the bias depends weakly on the model parameters.

For the models with $\lambda_V <{\cal O}(0.1)$ 
and the $\Lambda$CDM model we have $w^{\rm 2MASS}(\theta)>0$ 
for any angle $\theta$, so they can be 
compatible with the 2MASS data.
As we see in Fig.~\ref{fig2}, the model with 
$\lambda_V=1$ has a marginal 
positive ISW-galaxy cross-correlation. In this case the perturbation temporally 
enters the region ${\cal F}<0$, but the positive contribution to 
$C_l^{{\rm IG},{\rm 2MASS}}$ at high redshifts  leads to $w^{\rm 2MASS}(\theta)>0$. 
For $\lambda_V=10$, the minimum value of ${\cal F}$ is largely negative 
and hence $w^{\rm 2MASS}(\theta)<0$ for any angle $\theta$.
In Fig.~\ref{fig2}, we observe that the models with 
$\lambda_V \gtrsim 1$ are 
in tension with the 2MASS data. 
Thus, we have shown that the models with the large increase of $\Sigma$ at 
low redshifts [such as cases (c) and (d) in Fig.~\ref{fig1}] can be strongly
constrained from the ISW-galaxy cross-correlation data.

\section{Observational constraints}
\label{sec4}

In this section, we place observational constraints on the model 
given by the functions (\ref{G23}) by employing the ISW-galaxy cross-correlation data from 
the 2MASS and SDSS surveys \cite{ISWdata} as well as 
other observational data from CMB, BAO, SN Ia, $H(z)$, and RSDs. 
The latter data sets were also used in the likelihood analysis of 
Ref.~\cite{dFHT2017} to constrain the dark energy model in full GP 
theories, so we first briefly overview such a statistical method and then 
explore whether our dark energy model with $c_T^2=1$ can be 
compatible with all the data including  the ISW-galaxy 
cross-correlation.

\subsection{Priors on the model parameters}

The present dark energy model has the  following five free parameters:
\be
\Omega_{m0},~~h,~~s,~~p,~~\lambda_V\,.
\label{free}
\ee
At the background level, there are three free parameters, 
$\Omega_{m0}, h, s$, 
so we have only one additional quantity $s$ compared to the $\Lambda$CDM model.
At the level of perturbations, there are seven free parameters in full GP 
theories studied in Ref.~\cite{dFHT2017}.
Now, we consider the cubic-order GP theories with 
$c_T^2=1$, so this reduces the number of free parameters 
to six. Moreover, we consider the model with $q_V=1$, 
so we are left with the five parameters given by Eq.~(\ref{free}). 
We have chosen the parameter $\lambda_V$ instead of $\lambda$, 
as the former is directly related to the effect of intrinsic vector modes on $\mu$ and $\Sigma$.

As we mentioned in Sec.~\ref{ISWsub}, we set today's amplitude of overdensity $\sigma_{8}(0)$ to the Planck 
best-fit value. 
In the MCMC simulation, we also carried out 
the analysis by varying the initial value $\sigma_{8}(z_i)$ 
at ${\cal N}=-6$
in the $2\sigma$ range constrained by the Planck data 
in the $\Lambda$CDM model \cite{Planck2018}.  
The use of initial conditions 
$\sigma_{8}(z_i)$  that are the same as those in the $\Lambda$CDM 
model is plausible in that the evolution of perturbations 
in our model is very similar to that in the  
$\Lambda$CDM model during the deep matter era. 
We find that the resulting observational constraint on $\sigma_{8}(0)$ 
is similar to its $2\sigma$ Planck bound in the 
$\Lambda$CDM model. Moreover, 
the observational constraints on five parameters $\Omega_{m0}, h, s, p, \lambda_{V}$ are hardly affected by adding the parameter $\sigma_{8}(z_i)$ in the likelihood analysis.

In the MCMC simulation, we set the following priors on the 
parameter space of five model parameters.
\begin{itemize}
\item Today's density parameter of nonrelativistic matter:
$0.1 \leq \Omega_{m0} \leq 0.5$\,.

\item The normalized Hubble constant: 
$0.6 \leq h \leq 0.8$\,.

\item The deviation parameter from the $\Lambda$CDM 
model: $0 < s \leq 1$\,. 

\item The power $p$ in Eq.~(\ref{phipre}): 
$0 < p \leq 25$\,.

\item The parameter $\lambda_V$: 
$10^{-13} \leq \lambda_V \leq 15$\,.
\end{itemize}
In addition, we need to take into account the 
conditions for the absence of ghosts and Laplacian 
instabilities of scalar perturbations. 
They are given by 
\begin{itemize}
\item  $Q_{S} > 0$ and $c_{S}^{2} >0$ in the whole cosmological epoch. 

\item $0<ps \le 1$ to avoid the strong coupling at early times [see 
Eq.~(\ref{pscon1})].

\item $p (1 + s) \geq 1$ for avoiding the divergence of $c_{S}^{2}$ at early times 
[see Eq.~(\ref{pscon2})].
\end{itemize}
The other model parameters are known from the five parameters
in Eq.~(\ref{free}), say, $p_2=sp$ and $p_3=[p(1+2s)-1]/2$.

\subsection{Observational data}

We briefly explain the likelihood method and observational data used in our 
MCMC analysis. For more details, we refer the readers 
to Ref.~\cite{dFHT2017}.

\subsubsection{CMB}

To constrain the model from the CMB data, 
we resort to the following 
two CMB shift parameters:
\be
l_{a} = \frac{\pi \chi (z_{\ast})}{r_{\rm s}(z_{\ast})} \,,\qquad 
{\cal R} = \sqrt{\Omega_{m0}} H_{0} \chi(z_{\ast}) \,,
\ee
where $\chi(z)=\int_0^z H^{-1}(\tilde{z})d\tilde{z}$ is the comoving distance, 
and $r_{\rm s}(z)=\int_z^{\infty}c_{\rm s} H^{-1}(\tilde{z})d\tilde{z}$ is 
the comoving sound horizon with 
$c_{\rm s}=[3\{1+3\rho_{b 0}/(4\rho_{\gamma0})(1+z)^{-1}\}]^{-1/2}$
($\rho_{b 0}$ and $\rho_{\gamma0}$ are today's densities 
of baryons and photons, respectively). 
In the following, we fix today's baryon density parameter $\Omega_{b0}=\rho_{b0}/(3M_{\rm pl}^2 H_0^2)$ 
to the Planck best-fit value 
$\Omega_{b0}=0.02226$ \cite{Planck2015_14}.
For the decoupling redshift $z_*$, we employ the fitting formula of 
Hu and Sugiyama \cite{Sugiyama}.

The mean values of CMB shift  parameters constrained from the Planck 2015 
data are $\langle l_{a} \rangle = 301.77$ and 
$\langle \mathcal{R} \rangle = 1.4782$  with the deviations 
$\sigma(l_a) = 0.090$ and $\sigma(\mathcal{R}) = 0.0048$, respectively
\cite{Wang:2015tua,Planck2015_14}.
The $\chi^{2}$ statistics for these parameters is defined by
\be
\chi^{2}_{\rm CMB} = \bm{V}^{T} \bm{C}^{-1} \bm{V} \,,
\ee
where $\bm{V}^{T} \equiv ((l_{a} - \langle l_{a} \rangle)/\sigma(l_{a}),\, (\mathcal{R} - \langle \mathcal{R} \rangle)/\sigma(\mathcal{R}))$, 
and $\bm{C}^{-1}$ is the inverse of the normalized covariance matrix $\bm{C}$.
The components of $\bm{C}$ are given by 
$C_{11} = C_{22} = 1$ and $C_{12} = C_{12} = 0.3996$.

\subsubsection{BAO}

The observable associated with the BAO measurements is the ratio 
$r_{\rm BAO}(z_j) \equiv r_{\rm s}(z_d)/D_V(z_j)$
between the sound horizon $r_{\rm s}(z_d)$ at the redshift 
$z_d$ where baryons are 
released from the Compton drag of photons and the dilation scale $D_V(z_j)$ 
at the observed redshifts $z_j$. For the drag redshift $z_d$, 
we use the fitting formula of Eisenstein and Hu \cite{EHu1}.
The dilation scale is defined by
\be
D_{V}(z)=[z (1+z)^{2} D_{A}^{2}(z) H^{-1}(z)]^{1/3}\,,
\ee
where $D_A(z)=(1+z)^{-1} \int_0^z H^{-1} (\tilde{z})d \tilde{z}$ is the angular diameter distance.
For given $N$ data of $r_{\rm BAO}(z_j)$ with the error 
$\sigma(z_j)$, the $\chi^{2}$ estimator in BAO measurements is
given by 
\be
\chi^{2}_{\rm BAO} =\sum_{j=1}^{N} 
\frac{[r_{\rm BAO}(z_j)-\langle r_{\rm BAO}(z_j)\rangle]^2}
{\sigma^2(z_j)}\,,
\ee
where $\langle r_{\rm BAO}(z_j)\rangle$ is the mean observed 
value of each data. We exploit the BAO data extracted 
from the surveys of 6dFGS \cite{6dFGS}, SDSS-MGS \cite{SDSS-MGS}, BOSS \cite{BOSS}, BOSS CMASS \cite{BOSS-CMASS}, and Wiggle Z \cite{Wiggle-Z}.

\subsubsection{SN Ia}

The SN Ia has a nearly constant absolute magnitude $M \simeq -19$ 
at the peak of brightness. 
The observed apparent magnitude $m$ of SN Ia is different from its 
absolute magnitude $M$, whose difference is quantified as 
\be
\mu(z) \equiv m(z) - M = 5 \log_{10} 
\left[ \frac{d_L(z)}{10\, {\rm pc}}\right] \,,
\ee
where $d_L(z)=(1+z) \int_0^z H^{-1} (\tilde{z})d \tilde{z}$ is 
the luminosity distance from the observer to the source at redshift $z$.
The $\chi^2$ estimator in SN Ia measurements is defined by 
\be
\chi^{2}_{\rm SNIa} =
\sum_{j=1}^{N}\frac{\left[ \mu(z_j) - \langle \mu_{\rm obs}(z_j) 
\rangle \right]^{2}}{\sigma^2(z_j)} \,,
\label{chisq_SNIa}
\ee
where $N$ is the number of data sets, and  
$\langle \mu_{\rm obs}(z_j) \rangle$ 
is the mean observed value 
of $\mu (z_j)$ with the error $\sigma (z_j)$. 
We use the Union 2.1 data sets \cite{Suzuki:2011hu} 
for the computation of $\chi^{2}_{\rm SNIa}$.

\subsubsection{Local measurements of the Hubble 
expansion rate}

The direct measurement of the Hubble constant from the observations 
of Cepheids places the bound $h=0.7324 \pm 0.0174$ \cite{Riess:2016jrr}.
In addition, the Hubble expansion rate $H(z)$ at redshift $z$ 
can be constrained from the 
measurement of the ratio $r_H(z) \equiv r_{\rm s}(z_d)/H^{-1}(z)$
in BAO measurements. We define the $\chi^2$ statistics
associated with the local measurements of $H$, as 
\be
\chi_H^2=\frac{(h - 0.7324)^{2}}{0.0174^{2}} 
+\sum_{j=1}^{3}\frac{[r_{H}(z_j)
-\langle r_{H} (z_j)\rangle]^2}
{\sigma^2(z_j)}\,,
\ee
where $\langle r_{H} (z_j)\rangle$ is the mean observed value 
of $r_H (z_{j})$ at redshift $z_j$ 
with the error $\sigma (z_j)$. 
We exploit the three data provided 
by the BOSS measurement \cite{BOSS}.

\subsubsection{RSDs}

The RSD measurement can constrain the following quantity:
\be
y(z) \equiv f(z) \sigma_{8}(z) \,,
\ee
where $f(z)\equiv \delta'/\delta$ is the linear growth rate of
the matter density contrast. 
To compute $y(z)$ in our model, we resort to Eq.~(\ref{eqdelta_quasi}) 
derived under the quasistatic approximation for perturbations 
deep inside the sound horizon. 
This equation can be expressed as
\be
\delta'' + \frac{1 + (3 + 4 s) \Omega_{\rm DE}}{2\, (1 + s\, \Omega_{\rm DE})} \delta' - \frac{3}{2}\, \mu\, (1 - \Omega_{\rm DE}) \delta =0 \,,
\label{deleq}
\ee
where $\mu$ is given by Eq.~(\ref{mu_model}) with the scalar 
propagation speed squared (\ref{cS2_model}). 
In the deep matter era ($\Omega_{\rm DE} \ll 1$), we have
$\mu \simeq 1$, so the evolution of $\delta$ is similar to that 
in the $\Lambda$CDM model. 
We express $\delta$ in Fourier space as Eq.~(\ref{delD}) 
and choose the initial conditions $D'=D=e^{{\cal N}_i}$ 
at ${\cal N}_i=-6$. Since the growth rate $D(z)$ 
is known after solving Eq.~(\ref{deleq}),  
we obtain $\sigma_{8}(z)=\sigma_8(0) D(z)/D_0$ 
and $y(z)$ by adopting the Planck best-fit value 
$\sigma_8(0)=0.811$ \cite{Planck2018}.

If there are $N$ data sets with the mean observed value 
$\langle y_{\rm obs}(z_j) \rangle$ 
and the error $\sigma (z_j)$, 
the $\chi^{2}$ estimator for RSD measurements is
defined as 
\be
\chi^{2}_{\rm RSD} =
\sum_{j=1}^{N}\frac{\left[ 
y(z_{j}) - \langle y_{\rm obs}(z_{j}) \rangle 
\right]^{2}}{\sigma^2 (z_j)} \,.
\label{chisq_RSD}
\ee
We use the observational data given in Refs.~\cite{Hudson:2012gt, Beutler:2012px, Howlett:2014opa, Percival:2004fs, 
Song:2008qt, Blake:2011rj, Zheng:2018kgq, delaTorre:2013rpa, Okumura:2015lvp} for the computation of 
$\chi^2_{\rm RSD}$.

\subsubsection{ISW-galaxy cross-correlations}

The observable quantity associated with the ISW-galaxy cross-correlation is given by Eq.~(\ref{w_def}).
Then, we define the corresponding $\chi^2$ estimator, as
\be
\chi^{2}_{\rm IG} =
\sum_{A}
\sum_{j=1}^{N}\frac{\left[ w^{A}
(\theta_{j}) - \langle w_{\rm obs}^{A}
(\theta_{j}) \rangle \right]^{2}}{(\sigma_{j}^{A})^{2}} \,,
\label{chisq_ISW}
\ee
where $N$ is the number of data sets, 
$\langle w_{\rm obs}^{A}
(\theta_{j}) \rangle$ is the mean observed value  
of $w^{A}(\theta_j)$ with the error $\sigma_{j}^{A}$ on the data, and the subscript ``$A$" stands for different 
galaxy surveys.
To calculate $w^A(\theta_j)$ theoretically, we utilize the 
cross-correlation power spectrum (\ref{C_IG_limber})
with the Planck 2018 best-fit values $n_s=0.9649$ 
and $\sigma_8(0)=0.811$. 
For each model parameter, the quantities $\bar{\delta}_H$ 
and $b_s^A$ in Eq.~(\ref{C_IG_limber}) are computed 
according to the formulas (\ref{delHre}) and (\ref{bses}), respectively. 
For the observational data of $\langle w_{\rm obs}^{A}
(\theta_{j}) \rangle$ and $\sigma_{j}^{A}$, we choose 
those of 2MASS and SDSS surveys given 
in Ref.~\cite{ISWdata}.

\subsection{Likelihood results}

We perform the MCMC sampling over the allowed 
five-dimensional parameter space and compute the 
following $\chi^{2}$ statistics: 
\be
\chi^{2} = \chi^{2}_{\rm CMB} + \chi^{2}_{\rm BAO} 
+ \chi^{2}_{\rm SNIa}+ \chi^{2}_{H} 
+ \chi^{2}_{\rm RSD} + \chi^{2}_{\rm IG} \,.      
\ee
The best-fit model corresponds to the case in which 
$\chi^{2}$ is minimized.

\begin{figure}[htbp]
\begin{center}
\hspace{-1.0cm}
\includegraphics[width=6.85in]{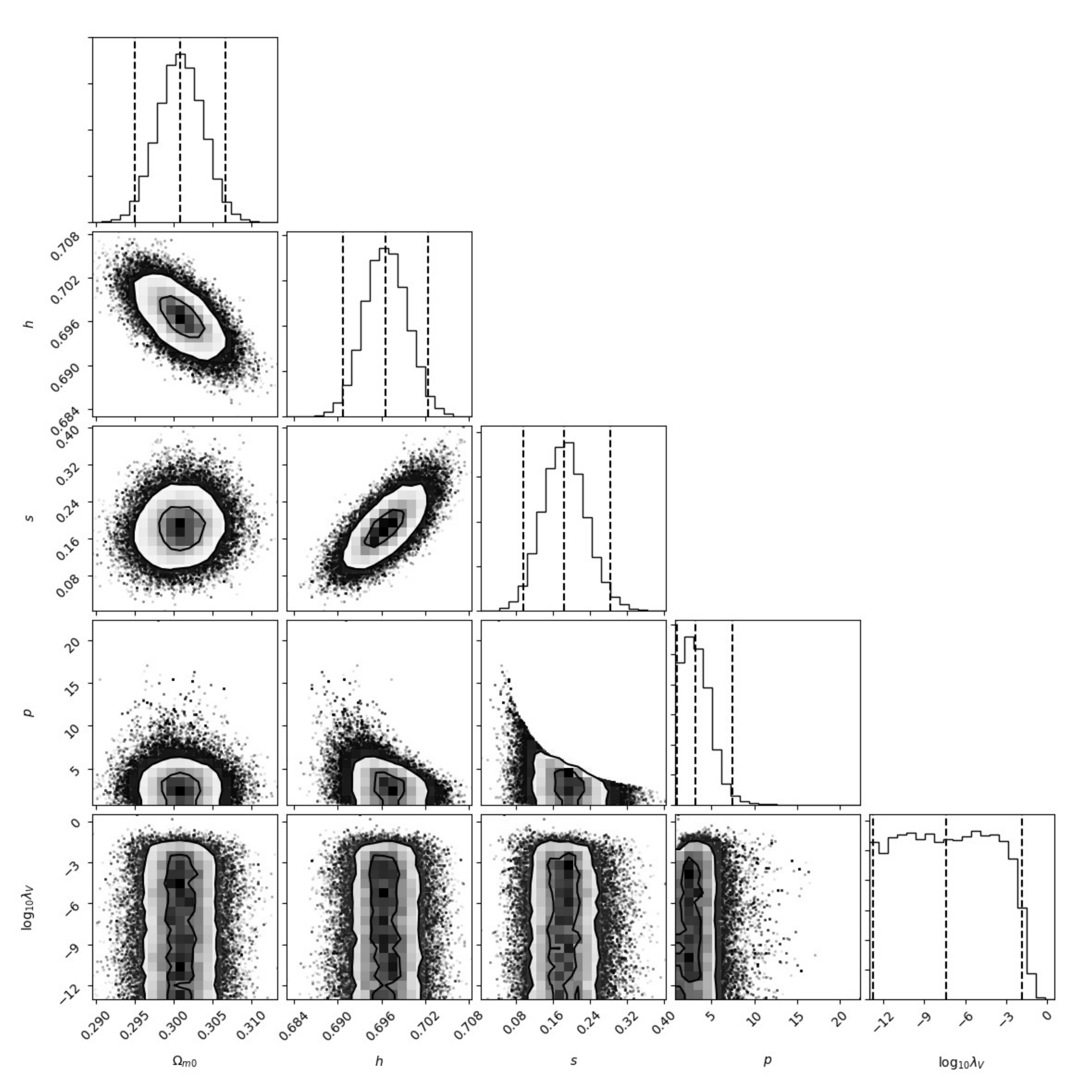}
\end{center}
\caption{\label{fig3}
Observational bounds on the five model parameters 
$\Omega_{m0},\,h,\,s,\,p,\,\lambda_V$ 
derived by the joint data analysis of CMB, BAO, SN Ia, 
$H_0$, RSDs, and the ISW-galaxy cross-correlation 
with the catalogues of 2MASS and SDSS. 
The vertical dashed lines represent the best fit (central) and 
the $2\sigma$ confidence limits (outside).
The quantities $\Omega_{m0}$, $h$, and $s$ 
are tightly constrained from the background expansion history.
The parameter $p$ is bounded from above from 
the theoretical prior $ps \le 1$. The quantity $\lambda_V$ 
is constrained to be $\lambda_V<0.015$ from the RSD 
and ISW-galaxy cross-correlation data. 
}
\end{figure}

In Fig.~\ref{fig3}, we show one-dimensional probability distributions for each parameter 
and two-dimensional observational contours for the combination of the 
five parameters (\ref{free}). The middle dashed lines in one-dimensional probability 
distributions represent the best-fit parameters. 
Considering the background expansion history alone with the data of 
CMB, BAO, SN Ia, and $H(z)$, there exists a global minimum of $\chi^{2}$ 
corresponding to the best-fit values of $\Omega_{m0}, h, s$. 
In the full MCMC analysis including the RSD and 
ISW-galaxy cross-correlation data,
the global minimum of $\chi^2$ is not uniquely fixed. 
There are several different sets of parameters giving similar lowest 
values of $\chi^{2}$, by reflecting the fact that the parameters 
$p$ and $\lambda_V$  are not well constrained from the data.
In other words, the models with some different sets of 
parameters lead to practically the same cosmological dynamics. 
One of the examples for such a set of model parameters is given 
by\footnote{The other examples of model parameters with 
$\chi_{\rm min}^2$ similar to Eq.~(\ref{chimin}) are 
$(\Omega_{m0},\, h,\, s,\, p,\, \lambda_{V}) = 
(0.3016,\, 0.696,\, 0.188,\, 4.541,\, 3.010 \times 10^{-3} )$
and 
$(0.3012,\, 0.696,\, 0.192,\, 3.602,\, 7.140 \times 10^{-12} )$. 
These values of $p$ and $\lambda_V$ are quite different from those 
in Eq.~(\ref{best-fit-params}).}
\be
\Omega_{m0} = 0.301, \quad
h = 0.697,\quad
s = 0.185, \quad
p = 3.078, \quad
\lambda_V =4.370 \times 10^{-8}\,,
\label{best-fit-params}
\ee
with the minimal value
\be
\chi_{\rm min}^{2} = 618.9\,.
\label{chimin}
\ee
The 2$\sigma$ bounds corresponding to these parameters are 
\ba
\Omega_{m0} = 0.301^{+0.006}_{-0.006},\quad 
h = 0.697^{+0.006}_{-0.006},\quad
s = 0.185^{+0.100}_{-0.089},\quad
p = 3.078^{+4.317}_{-2.119},\quad
\bar{\lambda}_V \leq \lambda_V < 0.015 \,,
\label{obbound}
\ea
where $\bar{\lambda}_V$ is the lower limit of the assumed prior. 

The observational bounds on $\Omega_{m0}$, $h$, and $s$ are similar to 
those derived in Ref.~\cite{dFHT2017} in full GP theories without the 
ISW-galaxy cross-correlation data. 
This means that the background expansion history mostly determines the
observational constraints on these three parameters. 
The model with $s=0$, i.e., the $\Lambda$CDM model, is outside the 
$2\sigma$ likelihood contour, so it is disfavored over the best-fit model  
with (\ref{chimin}) in cubic-order GP theories. 

We carry out the independent MCMC sampling for the $\Lambda$CDM 
model by varying the two parameters $\Omega_{m0}$ and $h$. 
We find that the best-fit $\Lambda$CDM model corresponds to 
$\Omega_{m0}=0.299$ and $h=0.687$ with $\chi^2_{\Lambda{\rm CDM}}=642.7$, 
whose $\chi^2$ is larger than (\ref{chimin}). 
In GP theories, the existence of the additional parameter $s$ to those 
in the $\Lambda$CDM model can reduce the tensions of the parameters 
$h$ and $\Omega_{m0}$ between CMB and low-redshift measurements.
In particular, the normalized Hubble constant $h$ shifts to the region 
between the best-fit values of CMB ($h \simeq 0.67$) \cite{Planck2018} 
and local measurements of $H_0$ ($h \simeq 0.73$) \cite{Riess:2016jrr}.

The observational contour in the two-dimensional 
$(p, s)$ plane of Fig.~\ref{fig3} is bounded by the prior 
$ps \le 1$ arising from the absence of the strong coupling 
problem of scalar perturbations in the asymptotic past. 
Compared to the $2\sigma$ upper limit $p<22.6$
derived in Ref.~\cite{dFHT2017} without imposing 
the prior $ps \le 1$, the upper bound on $p$  
is now reduced to $p<7.4$. 

As we see in the one-dimensional probability distribution of $\lambda_V$ 
in Fig.~\ref{fig3}, the central value of $\lambda_V$ is not well 
constrained from the data, but there exists the $2\sigma$ upper 
limit $\lambda_V<0.015$. 
In the limit that $\lambda_V \to 0$, we recover the values  
$\mu_{\rm dS}=\Sigma_{\rm dS}=1$ in GR. 
On using the best-fit parameters $s=0.185$ and $p=3.078$ 
with the bound $\lambda_V<0.015$, we obtain the limit
$\mu_{\rm dS}=\Sigma_{\rm dS}<1.011$ from Eq.~(\ref{muds}).
Thus, we have shown that the existence of the intrinsic vector 
mode can give rise to the values of $\mu$ and $\Sigma$ close 
to those in GR.  This behavior does not occur in 
scalar-tensor theories, as they correspond to the
other limit $\lambda_V \to \infty$.

\begin{figure}[ht]
\begin{center}
\hspace{-1.0cm}
\includegraphics[height=3.2in,width=3.3in,]{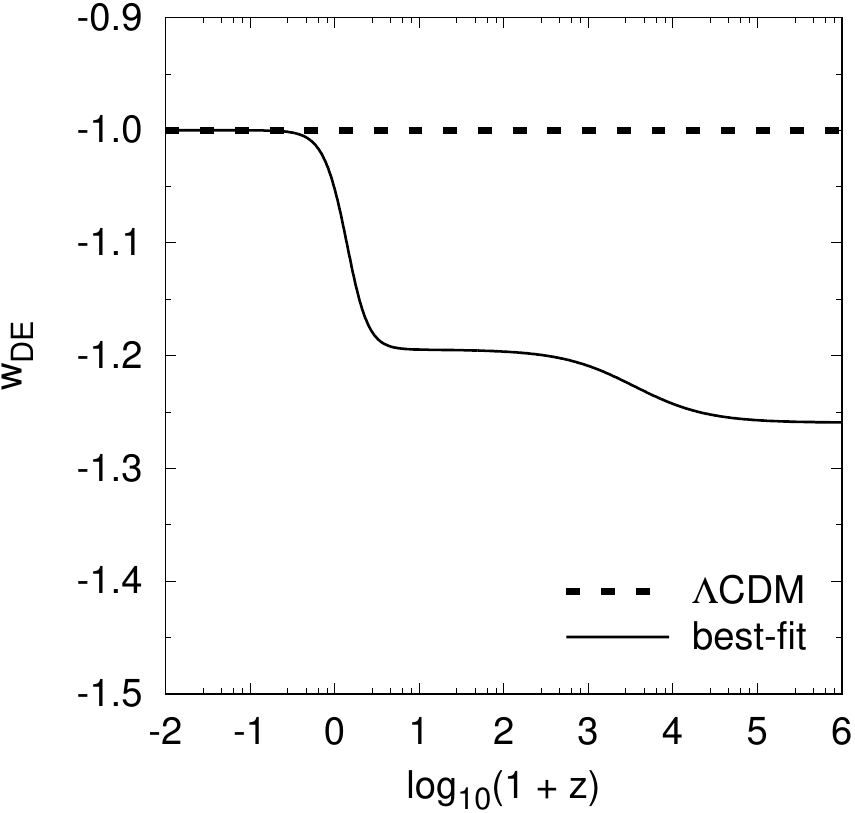}
\hspace{0.3cm}
\includegraphics[height=3.2in,width=3.3in]{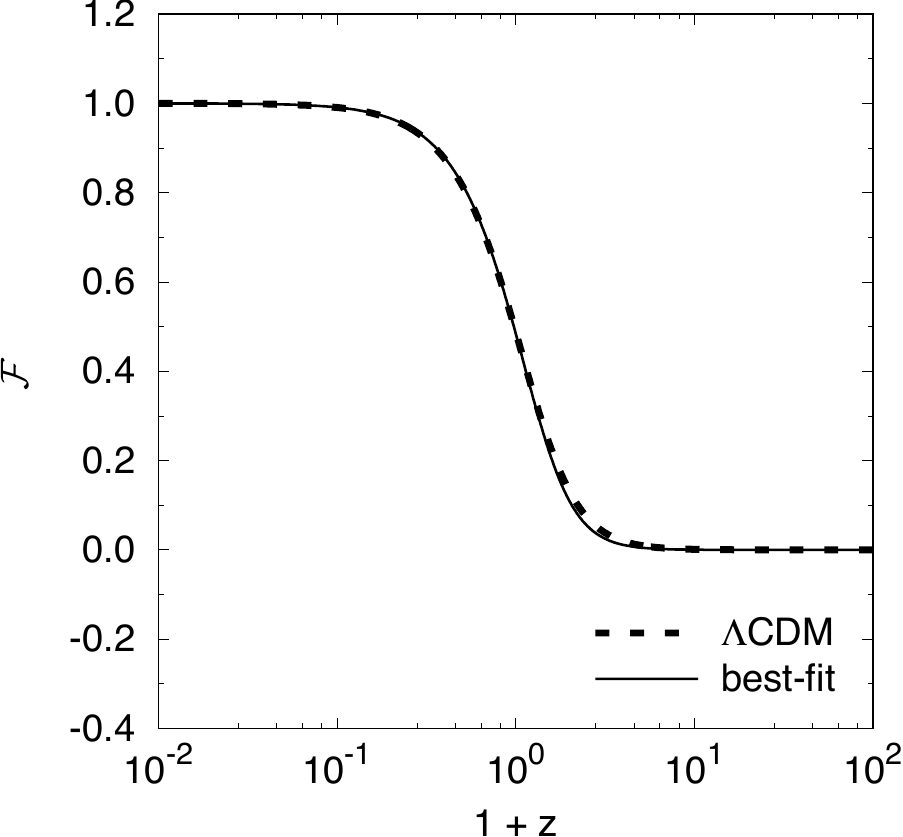}
\end{center}
\caption{\label{fig4}
(Left) Evolution of $w_{\rm DE}$ versus $1+z$ for 
the best-fit model parameters given by 
Eq.~(\ref{best-fit-params}) (solid line)
and for the $\Lambda$CDM model (dashed line).
(Right) Evolution of the quantity $\mathcal{F}$ 
defined by Eq.~(\ref{Fdef}) for the two models 
corresponding to the left panel. 
The background dynamics of the best-fit model in GP theories 
is different from that in the $\Lambda$CDM model, 
while the dynamics of perturbations is similar to each other.
}
\end{figure}

\begin{figure}[htbp]
\begin{center} 
\includegraphics[width=3.3in]{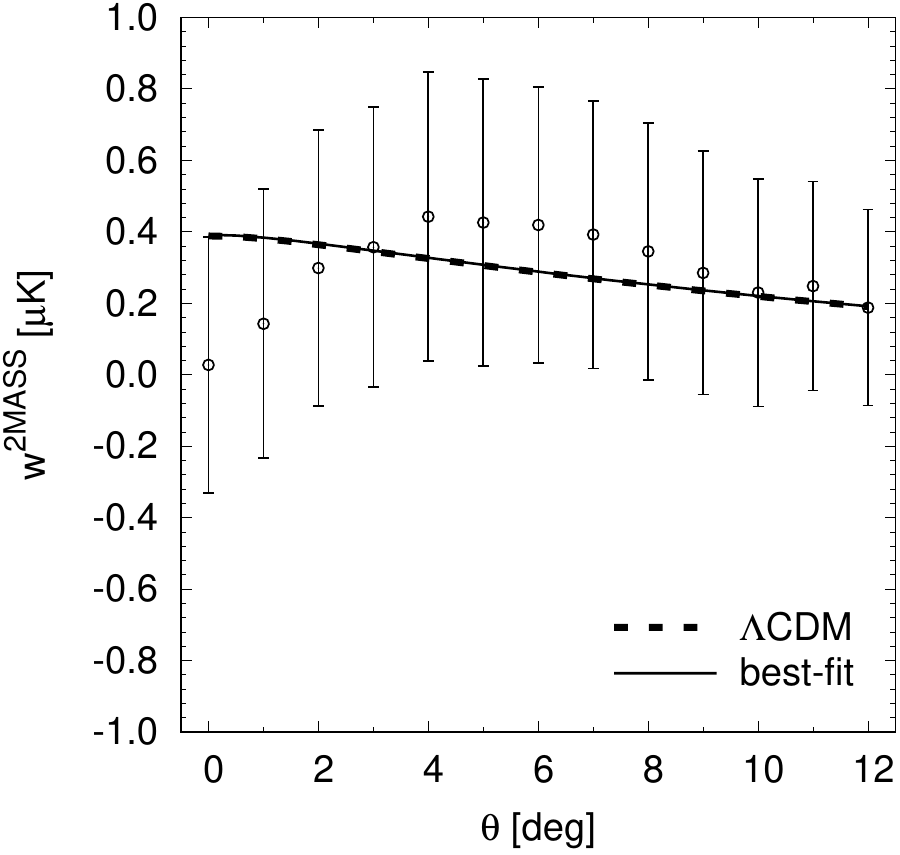}
\hspace{0.3cm}
\includegraphics[width=3.3in]{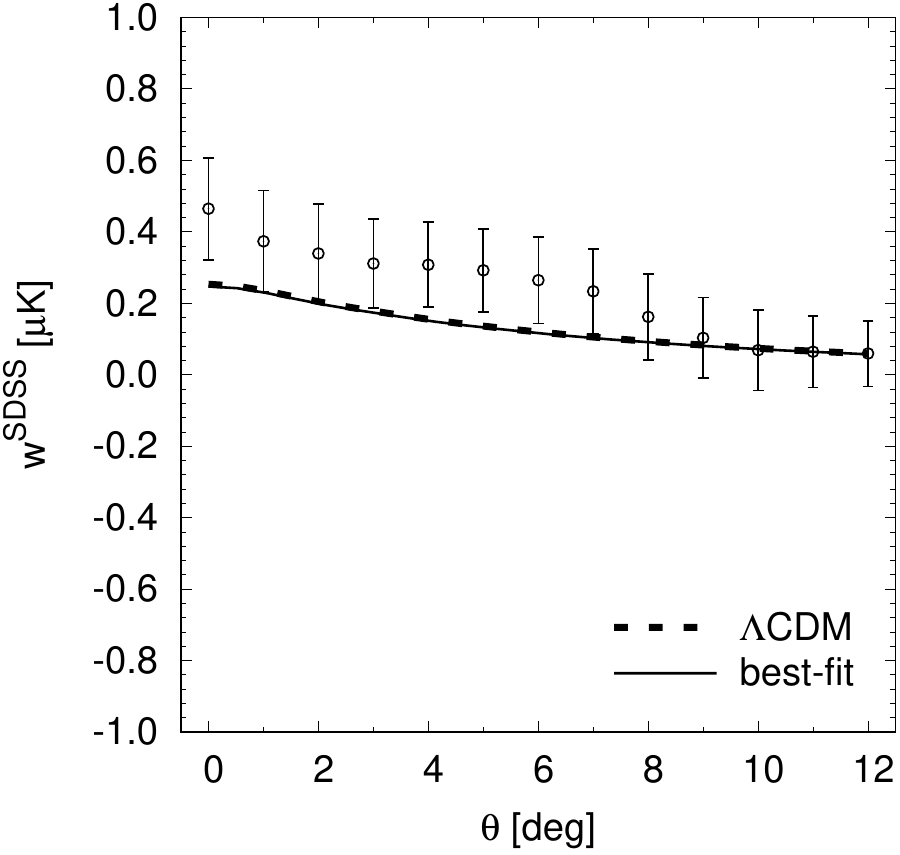}
\hspace{0.5cm} 
\end{center} 
\caption{\label{fig5}
The ISW-galaxy cross-correlation observable $w^A$ versus $\theta$
for the 2MASS (left) and SDSS (right) surveys for the best-fit model  
with the parameters (\ref{best-fit-params}) (solid line) and 
for the best-fit $\Lambda$CDM model (dashed line).
We also show the observational data with error bars 
in each galaxy survey. 
The cross-correlations predicted by the two models are 
almost the same as each other.
}
\end{figure}

We discuss the dynamics of background and perturbations 
for the best-fit model given by the parameters (\ref{best-fit-params}). 
As we see in the left panel of Fig.~\ref{fig4}, the best-fit model 
has the dark energy equation of state  
$w_{\rm DE}=-1.185$ during the matter era, 
which is followed by the approach to the de Sitter 
attractor ($w_{\rm DE}=-1$).
This is in stark contrast to the $\Lambda$CDM model in which 
$w_{\rm DE}$ is always equivalent to $-1$. 
On the other hand, in the right panel of Fig.~\ref{fig4}, we find that 
the evolution of the quantity ${\cal F}$, which appears in 
the ISW-galaxy cross-correlation spectrum $C_l^{\rm IG}$, is almost 
identical to that in 
the $\Lambda$CDM model. 
Indeed, substituting the best-fit values (\ref{best-fit-params}) 
into Eq.~(\ref{muds}), we obtain $\mu_{\rm dS}-1=\Sigma_{\rm dS}-1
=3.3 \times 10^{-8}$ and hence both $\mu$ and $\Sigma$ are 
very close to 1 throughout the cosmic expansion history. 

In Fig.~\ref{fig5}, we plot the ISW-galaxy cross-correlation observable $w^A(\theta)$ associated with two galaxy surveys 
for the best-fit model parameters (\ref{best-fit-params}).
Again, the theoretical curve in this model, which has the 
positive cross-correlation, is similar to that in 
the best-fit $\Lambda$CDM model.  
As we see in the left panel of Fig.~\ref{fig5}, 
the best-fit model can fit the 2MASS data quite well.
In the SDSS case, the model does not exhibit good fits 
to the data for
$\theta<7$ degrees.
To increase the values of $w^{\rm SDSS}(\theta)$ for the 
compatibility with the data, we 
require that the quantity $\Sigma$ is smaller than 1. 
However, this is not possible for cubic-order GP theories 
in which $\Sigma>1$ under the absence of ghosts and 
Laplacian instabilities. 
Then, the MCMC likelihood analysis finds the minimum value of 
$\chi^2$ with $\Sigma$ very close to 1.
In Fig.~\ref{fig2}, we observe that the model with $\lambda_V=0.1$ 
looks consistent with the 2MASS ISW-galaxy cross-correlation data.
However, the fact that this model is outside the $2\sigma$ limit 
$\lambda_V<0.015$ means that it is still in tension with the 
SDSS ISW-galaxy cross-correlation data.

The RSD measurements provide constraints on the dimensionless 
gravitational coupling $\mu$, 
which is the same as $\Sigma$ in cubic-order GP theories.
The RSD data \cite{Hudson:2012gt, Beutler:2012px, Howlett:2014opa, Percival:2004fs, 
Song:2008qt, Blake:2011rj, Zheng:2018kgq, delaTorre:2013rpa, Okumura:2015lvp} tend to favor the cosmic growth rate similar to that in GR or even smaller. 
Hence the models with $\mu$ close to 1 
are also favored from the RSD measurements.  
We performed the MCMC simulation without using the ISW-galaxy 
cross-correlation data and obtained the $2\sigma$ limit 
$\lambda_V<0.029$. Since this is weaker than the bound 
$\lambda_V<0.015$ derived by the full likelihood analysis, 
the ISW-galaxy data provide a more stringent bound 
on $\lambda_V$ than that constrained from the RSD data.

For the best-fit model parameters, the powers in the functions 
$G_2$ and $G_3$ are given by $p_2=0.4$ and $p_3=1.0$.
In this case, the coupling $G_3=b_3 X^{p_3}$ corresponds to 
that in the cubic vector Galileon. 
In scalar-tensor theories, if the cubic Galileon gives the 
dominant contribution to the dark energy density, 
this  leads to the negative ISW-galaxy cross-correlation incompatible with the observational 
data \cite{Kobayashi:2009wr,Kimura:2011td,Renk}. 
In GP theories, the 
existence of intrinsic vector modes can make the cubic 
vector Galileon compatible with the 
ISW-galaxy cross-correlation data by reducing 
$\Sigma$ to a value close to 1.
Thus, the dark energy model in GP theories can be 
observationally
distinguished from the corresponding counterpart in scalar-tensor theories. 

While $\chi_{\rm min}^2=618.9$ is smaller than 
$\chi_{\Lambda {\rm CDM}}^2=642.7$, our model has 
more free parameters than those in the $\Lambda$CDM model. 
To make comparison with these two models by taking into account 
the number of degrees of freedom, we resort to the Akaike information 
criterion (AIC) \cite{AIC} and Bayesian information criterion 
(BIC) \cite{BIC}. 
They are defined, respectively, by 
\be
{\rm AIC} = \chi^{2} + 2 \mathcal{P}\,, \qquad
{\rm BIC} = \chi^{2} + \mathcal{P} \ln(N_{\rm data})\,,
\ee
where $\mathcal{P}$ is the number of model parameters, and  
$N_{\rm data}$ is the number of data points.
For the best-fit model parameters (\ref{best-fit-params}),  
we obtain ${\rm AIC}=628.9$ and ${\rm BIC}=651.2$.
They are smaller than their best-fit values in the 
$\Lambda$CDM model: ${\rm AIC}= 646.7$ and ${\rm BIC}=655.6$.
Thus, even with the AIC and BIC, our model is 
statistically favored over the $\Lambda$CDM model.

\section{Conclusion}
\label{sec5}

In this paper, we placed observational constraints on a class 
of dark energy models in the framework of GP theories. 
{}From the GW170817 event, the speed of gravitational waves $c_T$ 
needs to be very close to 1 at the redshift $z<0.009$. 
Demanding that $c_T=1$ in GP theories, 
the allowed Lagrangians are up to cubic-order derivative 
interactions plus intrinsic vector modes.
Unlike the previous work \cite{dFHT2017}, we focus on 
the dark energy model satisfying the condition $c_T=1$ 
and included the ISW-galaxy cross-correlation data in the 
MCMC analysis to constrain the model further.

In scalar-tensor theories with the derivative coupling (including the Galileon), 
it is known that the dominance of cubic 
derivative couplings in the late Universe typically 
leads to the negative ISW-galaxy cross-correlation incompatible with observations. 
Since the same derivative coupling arises by 
taking the scalar limit $A_{\mu} \to \nabla_{\mu} \varphi$ 
in GP theories, one may anticipate that a similar property 
persists in cubic-order GP theories. In GP theories, 
however, there exist intrinsic vector modes associated with 
the transverse vector propagating degrees of freedom. 
Since the evolution of scalar perturbations on the FLRW 
background is affected by intrinsic vector modes, 
the observational predictions in GP theories are generally 
different from those in scalar-tensor theories.

In cubic-order GP theories, the dimensionless gravitational 
couplings $\mu$ and $\Sigma$, which are felt by matter and 
light, respectively, are given by 
$\mu=\Sigma=1+(\phi^2 G_{3,X})^2/(q_S c_S^2)$ 
under the quasistatic approximation. 
Provided that neither ghosts nor Laplacian 
instabilities of scalar perturbations are present 
($q_S>0$ and $c_S^2>0$), the gravitational interactions 
are enhanced ($\mu=\Sigma>1$) compared to those in GR. 
The effect of intrinsic vector modes on $\mu$ and $\Sigma$ 
arises through the quantity $\lambda_V$ defined 
by Eq.~(\ref{lambdaV}), where $q_V=1$ for the model (\ref{G23}).
This allows the possibility for realizing the values of 
$\mu$ and $\Sigma$ close to 1.

In Sec.~\ref{ISWsub}, we provided a general formula for 
the ISW-galaxy cross-correlation spectrum $C_l^{\rm IG}$ 
for the scale-independent growth of linear perturbations. 
A key quantity characterizing the sign of $C_l^{\rm IG}$ 
is the factor ${\cal F}=1-(\ln D\Sigma)'$ 
[see Eq.~(\ref{C_IG_limber})]. The necessary condition for 
the negative cross-correlation to occur is that 
the perturbation enters the region ${\cal F}<0$ at low 
redshifts. In Sec.~\ref{sec3C}, we studied the evolution of 
${\cal F}$ for the concrete dark energy model 
(\ref{G23}) and computed the ISW-galaxy cross-correlation 
observable $w^A (\theta)$ corresponding to 
the 2MASS galaxy survey.
As the quantity $\lambda_V$ decreases, 
the gravitational couplings (\ref{muds}) on the de 
Sitter solution approach the values 
$\mu_{\rm dS}=\Sigma_{\rm dS}=1$,
so the model exhibits a better compatibility 
with the observational data (see Fig.~\ref{fig2}).

In Sec.~\ref{sec4}, we performed the MCMC analysis for 
the dark energy model (\ref{G23}) in cubic-order GP 
theories by using the ISW-galaxy cross-correlation data 
of the 2MASS and SDSS surveys combined with the CMB, 
BAO, SN Ia, $H(z)$, and RSD data. 
The evolution of the dark energy equation of state
during the matter era is given by $w_{\rm DE}=-1-s$, 
where $s$ is a positive constant. 
The parameter $s$ is constrained to be 
$s = 0.185^{+0.100}_{-0.089}$ at 95 \%\,CL, 
so the model with $s>0$ is favored over the 
$\Lambda$CDM model ($s=0$).
At the background level, this property is attributed to 
the fact that the presence of the additional parameter 
$s$ to $H_0$ and $\Omega_{m0}$ 
can reduce the tensions of $H_0$
between CMB and low-redshift measurements.

For the cosmic growth history, the model can be compatible 
with both the ISW-galaxy cross-correlation and 
RSD data thanks to the existence of intrinsic vector modes.
{}From the MCMC simulation, we derived 
the $2\sigma$ bound $\lambda_V<0.015$. 
The likelihood analysis without the ISW-galaxy 
cross-correlation data placed the $2\sigma$ 
constraint $\lambda_V<0.029$. 
This means that inclusion of the ISW-galaxy data, 
in particular, the SDSS data, provides a tighter constraint 
on $\lambda_V$ compared to that obtained from the RSD data. 
The existence of intrinsic vector modes can make the model  
compatible with both the ISW-galaxy 
cross-correlation and RSD data by reducing 
$\mu$ and $\Sigma$ close to 1.
As we see in Figs.~\ref{fig4} and \ref{fig5}, 
the evolution of $w_{\rm DE}$ in the best-fit case 
is clearly different from that in the $\Lambda$CDM 
model, while the evolution of perturbations 
is similar to each other.
 
We have thus shown that the dark energy model in cubic-order 
GP theories exhibits the interesting feature of fitting the observational 
data better than the $\Lambda$CDM model. 
We would like to stress that not only the best-fit $\chi^2$ but also 
the AIC and BIC in our model are smaller than those in the 
$\Lambda$CDM model.
Since the scalar-tensor 
analogue of GP theories corresponds to the limit 
$\lambda_V \to \infty$, this nice property does not hold in 
Horndeski theories with the dominance of cubic 
derivative couplings for the late-time cosmological dynamics. 
It remains to be seen how the future high-precision 
observational data constrain the dark energy model 
in GP theories further.

\section*{Acknowledgements}

S.N. and S.T. thank warm hospitalities to the members in YITP where
part of this work was done. A.D.F. thanks Tommaso Giannantonio, Gongbo
Zhao, and Jinglan Zheng for useful comments. A.D.F. was supported by JSPS
KAKENHI Grant No.~16K05348.  R.K. is supported by the Grant-in-Aid for
Young Scientists B of the JSPS No.\,17K14297.  S.T. is supported by the
Grant-in-Aid for Scientific Research Fund of the JSPS No.~16K05359 and
MEXT KAKENHI Grant-in-Aid for Scientific Research on Innovative Areas
``Cosmic Acceleration'' (No.\,15H05890).


\end{document}